%% file: main.tex
\documentclass[sigconf,9pt,nonacm]{acmart}
\usepackage{tabularx}
\usepackage{multirow}
\usepackage{colortbl}
\usepackage{pifont}
\usepackage{xspace}

\newcommand{\emptymark}{{%
  \setlength{\unitlength}{0.75ex}\linethickness{0.07ex}%
  \begin{picture}(2,2)\put(1,1){\circle{2}}\end{picture}}}
\newcommand{\halfmark}{{%
  \setlength{\unitlength}{0.75ex}\linethickness{0.07ex}%
  \begin{picture}(2,2)
    \put(1,1){\circle*{2}}
    \put(0,0){\textcolor{white}{\rule{1\unitlength}{2\unitlength}}}
    \put(1,1){\circle{2}}
  \end{picture}}}
\newcommand{\fullmark}{{%
  \setlength{\unitlength}{0.75ex}%
  \begin{picture}(2,2)\put(1,1){\circle*{2}}\end{picture}}}
\usepackage{xurl}
\usepackage{placeins}
\usepackage{float}
\usepackage{stfloats}
\newcommand{\sysname}{\texttt{RouterWeave}\xspace}
\setcopyright{none}
\acmDOI{}
\acmISBN{}
\title{\sysname{}: A Unified Observation Layer for Cross-Platform Activity Analysis via Home Routers}
\newcommand{\affilmark}{\texorpdfstring{\textsuperscript{\textdagger}}{}}
\newcommand{\equalcontribmark}{\texorpdfstring{\textsuperscript{*}}{}}
\author{Chengeng Liu\equalcontribmark}
\author{Ronghua Li\equalcontribmark}
\author{Yan Long\affilmark}
\author{Shinan Liu}

\makeatletter
\renewcommand{\@mkauthors}{%
  \global\setbox\mktitle@bx=\vbox{%
    \unvbox\mktitle@bx\medskip\centering
    {\@authorfont\let\and\quad\authors\par}
    {\normalsize The University of Hong Kong\qquad
     \affilmark{}Hong Kong University of Science and Technology (Guangzhou)\par}
    \bigskip}}
\makeatother
\begin{document}
\begin{abstract}
Smart homes bring together devices from different operating systems and vendor ecosystems, but each platform reports activity only for the devices and features it supports, leaving gaps in household-wide activity monitoring.
Interoperability standards such as Matter and software integrations connect and control supported devices, yet vendor restrictions and licensing barriers limit access to their data, while hardware constraints hinder broader integration.
Our key insight is that encrypted network traffic provides complementary evidence of activity across platform boundaries, and the existing home router offers a shared point for collecting and analyzing this evidence.
Accordingly, we present \sysname{}, a router-side observation framework that combines cross-platform traffic collection, task-adaptive processing, and unified activity records for household applications.
These records preserve activity meaning, temporal extent, and source-device attribution while distinguishing discrete operations from sustained activity intervals.
We evaluate \sysname{} across four datasets, native execution on a commodity home router, and two novel case studies: cross-platform Screen Time monitoring and appliance energy estimation.
\sysname{} achieves 94.41\% macro-F1 on application activities and 97.46\% on IoT operations, outperforming baselines by up to 20.99 percentage points, while using under 7~MB of router memory per task and supporting 54 concurrent devices within a 5-second update window.
In end-to-end case studies, \sysname{} reconstructs 600 Screen Time timelines across Android, iOS, and Windows with 92.15\% accuracy (24.84-second mean active-time error over 19-minute averages) and recovers 49 of 55 appliance operations over four days, tracking ON/OFF states with 92.84\% agreement and estimating electricity use within a 4.28\% error. Together, these results establish the commodity home router as practical infrastructure for whole-home activity sensing.
The code is open sourced at \url{https://github.com/naisslab/routerweave}.
\end{abstract}
\maketitle
{\renewcommand{\thefootnote}{\fnsymbol{footnote}}%
 \footnotetext[1]{Both authors contributed equally to this work.}}
\pagestyle{plain}
\raggedbottom
\section{Introduction}
Smart homes bring together a wide range of devices, from phones, laptops, and tablets to televisions, appliances, and sensors~\cite{kumar2019allthings}.
Users may select these devices from different vendors to suit their needs, budgets, and preferences.
Beyond hardware, these products also come with their own software features and services, shaped by the ecosystems behind them. For households, this variety provides greater flexibility in adopting features of devices across platforms.

This diversity, however, makes it harder to obtain a coherent view of activities across different platforms. \textit{A resident may move between an iPhone and a Windows laptop while using appliances and sensors chosen from several vendors, so everyday activities often extend across platforms by construction}. For example, recognizing a person's return home for safety or assisted living can require correlating events from a smart lock, contact sensor, doorbell, lights, and camera supplied by different vendors~\cite{wan2022iotmosaic}. Likewise, whole-home energy management depends on coordinating devices and energy services across vendors~\cite{csaMatter14}. To this end, applications that rely on an understanding of everyday life call for \textbf{\emph{cross-platform observation}}, a coherent account of activity across operating systems, device categories, and vendor ecosystems.

Existing solutions provide cross-device activity views through native ecosystems, interoperability standards~\cite{csaMatter14}, integration platforms~\cite{homeassistantEntities,homeassistantMatter,homeassistantMyQ,openhabConcepts}, or endpoint monitoring~\cite{qustodioPlatforms,activitywatchDocs}. Although effective within their supported domains, these approaches impose various requirements on \textbf{(i) protocol support}: proprietary platforms such as Apple and Google typically have their own integration protocols; \textbf{(ii) certified information access}: information that can be accessed is usually constrained and platform-dependent; and \textbf{(iii) multi-endpoint}: cross-platform and device information gathering rely on multiple interfaces rather than a unified endpoint. Accordingly, their coverage remains bounded by the devices and activities exposed by each standard or integration, as shown in Table~\ref{tab:integration-routes}. For cross-platform applications, the challenge is thus not merely reconciling heterogeneous data schemas, but accommodating devices with diverse protocols and policies through a consistent observation mechanism. So far, no single integration route can be assumed to accommodate every smart home device from a single endpoint.

\begin{table*}[t!]
\caption{Device coverage and deployment requirements of household activity observation approaches. \fullmark{} Yes; \emptymark{} No; \halfmark{} Depends on the product, integration, or existing device support.}
\label{tab:integration-routes}
\centering
\small
\setlength{\tabcolsep}{2.5pt}
\renewcommand{\arraystretch}{1.0}
\begin{tabularx}{\linewidth}{@{}>{\raggedright\arraybackslash}p{0.20\linewidth} >{\raggedright\arraybackslash}p{0.24\linewidth} >{\raggedright\arraybackslash}X *{4}{c}@{}}
\toprule
Approach & Examples & Device coverage & {\scriptsize\shortstack[b]{No Proprietary\\Protocol}} & {\scriptsize\shortstack[b]{No Hardware/\\Software Changes}} & {\scriptsize\shortstack[b]{No Licensing/\\Membership}} & {\scriptsize\shortstack[b]{Single-point\\Sensing}} \\
\midrule
Native ecosystems & Apple~\cite{appleHomeKitModel}, Google~\cite{googleHomeDataModel}, Microsoft~\cite{microsoftFamilySafety}, Samsung~\cite{samsungWellbeing} & Same-ecosystem devices & \emptymark & \halfmark & \halfmark & \emptymark \\
Interoperability standards & Matter~\cite{csaMatter14}, Zigbee~\cite{csaZigbee}, Z-Wave~\cite{zwaveAlliance} & Certified devices & \emptymark & \halfmark & \emptymark & \emptymark \\
\textit{Third-party tools} & & & & & & \\
\hspace*{0.9em}Monitoring software & Qustodio~\cite{qustodioPlatforms}, ActivityWatch~\cite{activitywatchDocs} & Supported phones and PCs & \fullmark & \emptymark & \halfmark & \emptymark \\
\hspace*{0.9em}Integration platforms & Home Assistant~\cite{homeassistantEntities}, openHAB~\cite{openhabConcepts} & Devices with integrations & \emptymark & \halfmark & \halfmark & \emptymark \\
\midrule
Passive network observation & \textbf{\sysname{} (this work)} & Network-visible devices & \fullmark & \fullmark & \fullmark & \fullmark \\
\bottomrule
\end{tabularx}
\end{table*}

In recent years, network traffic has offered a passive sensing potential for observing device activities across platform boundaries. For network-connected smart devices, their communication passes through a shared gateway regardless of the vendor or operating system, leaving behavioral evidence in packet timing, sizes, directions and service interactions even when payloads are encrypted. Prior studies have used this evidence to fingerprint websites~\cite{wang2014wf,panchenko2016wf}, recognize mobile applications~\cite{taylor2016appscanner,flowprint2020}, and infer IoT device events and behaviors~\cite{pingpong,homesnitch2019,liu2023amir,li2025wifinger}, but these efforts remain confined to individual web, mobile, or IoT domains and do not provide a shared activity representation across them. This motivates us to develop \textbf{network traffic as a passive and pervasive observation medium for cross-platform activity analysis}.

However, building such a unified cross-platform sensing layer with network traffic raises two challenges:
\begin{enumerate}
\item \textbf{Cross-platform heterogeneity.} Traffic and service characteristics vary across both device categories and platform implementations. Sustained web or media activities may involve extended flows and quiet periods, whereas IoT operations often produce short, intermittent bursts. The same activity can also exhibit different communication patterns and service associations on Android, iOS, and Windows. A common observation layer must accommodate these different temporal structures and interpret platform-dependent evidence without assuming one universal traffic signature.
\item \textbf{Practical deployment.} A unified observation layer needs a practical place to run. Reported online deployments use host-based processing, cloud middleboxes, or additional controllers~\cite{taylor2016appscanner,flowprint2020,recon,homesnitch2019}. Adding dedicated hardware can provide the required computation, but introduces installation and configuration effort and may not suit every household's network setup. Existing gateways and routers already provide a shared observation point, but their limited compute and memory must also support normal network operation. These constraints make it difficult to run traffic acquisition, inference, and activity reconstruction in real time on the same device. The deployment challenge is therefore to support continuous cross-platform observation on existing household infrastructure within its resource limits.
\end{enumerate}

In this work, we present \sysname{}, a router-side activity sensing framework that transforms the home router into an observation infrastructure for activities across heterogeneous smart-home devices.  Its design combines shared traffic acquisition, task-adaptive inference, and unified activity management to provide a coherent view of device activities.

\sysname{} separates the collection of network evidence from the methods used to interpret it. Because sustained activities and discrete operations differ in temporal structure, \sysname{} supports task-specific input construction, features, and models over the same traffic source. Despite these differences in recognition, \sysname{} organizes the resulting outputs into a shared activity history that preserves the temporal meanings of sustained activities and discrete operations and makes them available through common application interfaces. Platform-specific interfaces are used for activity annotation during development, while deployed inference relies on network observations alone.

Online observation introduces a related constraint: the temporal context used for inference determines how long data must be retained and when an estimate can be produced. \sysname{}'s online path processes incoming traffic using bounded history, coupling acquisition and inference without waiting for a completed recording. Model and temporal-processing choices are therefore considered together, accounting for recognition quality, memory use, computation, and output latency.

We evaluate \sysname{} across four datasets, native execution on a home router (GL-MT3000~\cite{glinetMT3000Specs}), and two case studies: cross-platform Screen Time monitoring for continuous activities and appliance energy estimation from bursty operations. \sysname{} outperforms baselines by up to 20.99 percentage points in macro-F1, using under 7~MB of router memory per task and supporting up to 54 concurrent devices (over twice the average household size) within a 5-second update window. For Screen Time, \sysname{} reconstructs 600 timelines (averaging 19 minutes) across Android, iOS, and Windows with 92.15\% accuracy and a 24.84-second mean active-time error ($\sim$2.2\%). Over four days of energy estimation, it recovers 49 of 55 operations, tracks ON/OFF states with 92.84\% agreement, and estimates total electricity use within a 4.28\% error. These experiments reveal two key insights: (1) cross-platform sensing requires matching traffic grouping and fusion to the activity type, as isolating IoT bursts from background traffic improves event F1 by 19.66 percentage points and combining traffic behavior with service context raises accuracy during quiet video buffering from 60.32\% to 96.83\%; and (2) while brief bursts can track long-term appliance states without constant polling, a single misread ON/OFF switch flips the tracked state for hours, showing that state tracking relies heavily on telling opposite actions apart.

To summarize, this paper makes three key contributions:
\begin{enumerate}
\item \textbf{To our knowledge, this is the first router-native system to transform encrypted network traffic into unified, application-ready activity histories across personal and IoT devices.} \sysname{} combines shared router-side traffic observation, task-adaptive recognition, and unified activity management to provide activity information across heterogeneous platforms through \mbox{common application interfaces.}

\item \textbf{We present two case studies demonstrating the applicability of the unified observation architecture.} Through cross-platform Screen Time monitoring and appliance energy estimation, we demonstrate how \sysname{} supports downstream tasks with distinct activity structures \mbox{and temporal requirements.}

\item \textbf{We open-source a large-scale cross-platform traffic dataset and its collection methodology.} We will release over 9,000 verified activity windows spanning Android, iOS, and Windows captures and over 5,000 annotated smart-home operations, together with the instrumented collection toolchain, so that the community can extend coverage to new platforms and households under the same provenance discipline.
\end{enumerate}

We will release the code and datasets to facilitate research in this direction for the community.
\section{Related Work}
This section reviews use cases that draw on activity information across devices (\S\ref{sec:rw-apps}), the information exposed by existing platforms and integration systems (\S\ref{sec:rw-platform}), and network-based approaches to activity inference (\S\ref{sec:rw-traffic}).

\begin{table*}[t!]
\caption{Limitations of existing traffic-analysis systems in representing a common activity layer across personal and IoT devices. Symbols denote no support (\emptymark), support in some methods within the category (\halfmark), and support (\fullmark).}
\label{tab:prior-works}
\centering
\small
\setlength{\tabcolsep}{3pt}
\renewcommand{\arraystretch}{1.1}
\begin{tabularx}{\linewidth}{@{}>{\raggedright\arraybackslash}p{0.20\linewidth} >{\raggedright\arraybackslash}X *{6}{c}@{}}
\toprule
& & \multicolumn{4}{c}{Activity coverage} & \multicolumn{2}{c}{Efficiency and feasibility} \\
\cmidrule(lr){3-6}\cmidrule(lr){7-8}
Category & Representative methods & Web & Mobile & IoT & \shortstack{Cross-\\platform} & Lightweight & Real-time \\
\midrule
Website fingerprinting & Wang et al.~\cite{wang2014wf}; Panchenko et al.~\cite{panchenko2016wf} & \fullmark & \emptymark & \emptymark & \emptymark & \emptymark & \emptymark \\
Mobile activity inference & AppScanner~\cite{taylor2016appscanner}; FlowPrint~\cite{flowprint2020}; ReCon~\cite{recon} & \emptymark & \fullmark & \emptymark & \emptymark & \halfmark & \fullmark \\
IoT activity and device inference & HomeSnitch~\cite{homesnitch2019}; PINGPONG~\cite{pingpong}; IoTAthena~\cite{iotathena}; CP-IoT~\cite{cpiot}; N-BaIoT~\cite{nbaiot2018}; WiFinger~\cite{li2025wifinger} & \emptymark & \emptymark & \fullmark & \emptymark & \halfmark & \halfmark \\
\midrule
Cross-platform observation & \textbf{\sysname{} (this work)} & \fullmark & \fullmark & \fullmark & \fullmark & \fullmark & \fullmark \\
\bottomrule
\end{tabularx}
\end{table*}

\subsection{Cross-Platform Smart Home Use Cases}
\label{sec:rw-apps}
Integrating information across devices has supported a wide range of smart-home use cases, including personal well-being, sustainability, and safety~\cite{lyngs2024distraction,saha2014energylens,wan2022iotmosaic}.

In personal well-being, digital self-control studies identify the importance of selecting interventions according to users' goals across their devices. Lyngs et al. help participants combine smartphone applications, browser extensions, and system settings around these goals, finding improved self-reported control over device use~\cite{lyngs2024distraction}. Cross-device sensing also supports health monitoring in everyday living environments: Capstick et al. combine room-motion and sleep-mat records to estimate urinary tract infection risk in people with dementia, finding that physiological and sleep measurements improve prediction over activity sensing alone~\cite{capstick2024monitoring}.

Moreover, sustainability systems use cross-device information to connect resource consumption with the activities and conditions that determine it. EnergyLens, for example, combines household electricity measurements with smartphone Wi-Fi and acoustic sensing to associate appliance use with occupants and rooms, improving activity recognition over electricity measurements alone~\cite{saha2014energylens}. In residential irrigation, ERIC combines local rainfall estimates from doorbell cameras with weather-station data to schedule sprinkler operation, with field-data analysis showing lower estimated over-irrigation than scheduling based on weather-station rainfall alone~\cite{liu2024eric}.

These examples illustrate a broader demand for access to information across platforms, enabling systems to interpret activity in context and coordinate responses to users' goals and changing household conditions.

\subsection{Cross-Platform Integration in Smart Homes}
\label{sec:rw-platform}
Proprietary ecosystems such as Apple Home and Google Home keep device data and modeling within their own walls~\cite{appleHomeKitModel,googleHomeDataModel}, making it difficult to achieve cross-platform unification. Matter standardizes an application-layer data model, device types and clusters~\cite{matterDataModel}, so that certified devices expose a common interface to every participating platform. Nonetheless, bringing a device onto that model falls to vendors and platform developers. As a result, product-specific capabilities must be adapted into standard clusters, and those not adapted remain unavailable. Integration platforms take a different route. Instead of forcing devices to use one data model, Home Assistant and openHAB adapt to each device through a common entity layer built from device and service-specific integrations~\cite{homeassistantEntities,openhabConcepts}. However, this means that for every new product family, there is a need to write and maintain another integration, and recorded history extends only as far as the underlying integration exposes states and events.

Cross-platform integration is not limited to appliances. Qustodio and ActivityWatch track application and web usage across phone and desktop operating systems through software installed on each device~\cite{qustodioPlatforms,activitywatchDocs}, offering a usage view that no single hardware vendor provides.

Although these systems differ in mechanism, they mostly converge on two limitations: scalability and representational complexity. Each system unifies only the states, events, or per-device usage that vendors choose to expose, and every new product requires additional integration work, whether in the form of cluster mappings, hand-written integrations, or installed clients. In addition, each system describes devices in its own schema, so no vendor-independent data model of device behavior exists, and no system tracks user activity continuously across the phones, laptops, and appliances that share the same smart home environment.

\subsection{Network Traffic as an Activity Observation Proxy}
\label{sec:rw-traffic}

Most devices in a smart home reach the internet through the same home network, so their traffic can be observed at a single point. Over the past decade, a growing body of work has used this traffic for a range of inference tasks, even when payloads are encrypted~\cite{papadogiannaki2021survey,guo2026pacc}.

Early efforts established the feasibility with classical machine learning. In website fingerprinting, k-NN and SVM classifiers identified visited pages from packet sizes and timings~\cite{panchenko2016wf,wang2014wf}. Similar approaches were carried over to mobile traffic: AppPrint~\cite{miskovic2015appprint} and AppScanner~\cite{taylor2016appscanner} classify applications from flow statistics, and FlowPrint~\cite{flowprint2020} derives semi-supervised fingerprints that work across Android and iOS. ReCon~\cite{recon} addresses a different target: it uses decision trees over request contents to detect personal-information leaks and supports user control through a VPN middlebox.

Later, deep learning emerged as another popular direction. Deep Fingerprinting~\cite{sirinam2018wf}, automated website fingerprinting~\cite{rimmer2018wf}, Var-CNN under data scarcity~\cite{varcnn2019}, and Tik-Tok~\cite{tiktok2020}, which relies on inter-packet timing alone, reported improvements in website-identification accuracy. On mobile traffic, deep classifiers such as MIMETIC~\cite{mimetic2019} and Deep Packet~\cite{deeppacket2020} report high per-flow accuracy, at the cost of substantially larger models.

Fingerprinting approaches target more intrinsic features of network activity. Rather than training general classifiers, these works derive efficient fingerprints from traffic and recognize events by matching packets and flows against known signatures. Most of them aim to identify what a device is doing, extracting event signatures that reveal individual IoT device events~\cite{pingpong,iotathena,li2025wifinger} or summarizing traffic into semantic behaviors~\cite{homesnitch2019}. Beyond event fingerprinting, related methods fingerprint device types~\cite{iotsentinel2017,sivanathan2019}, model device states~\cite{behaviot2023}, or flag compromised devices~\cite{nbaiot2018}.

Together, these studies establish that traffic metadata can reveal behavioral information despite payload encryption. However, as listed in Table~\ref{tab:prior-works}, three limitations prevent their direct application to cross-platform activity observation.

\noindent\textbf{(1) Activity coverage.} Existing studies typically examine mobile application usage, web browsing, and IoT device events in isolation, using domain-specific label spaces, traffic captures, and model families. Their recognition results therefore do not provide a shared activity layer spanning these domains.

\noindent\textbf{(2) Evidence integration.} Existing approaches exploit traffic statistics, packet-level representations, and characteristic signatures, but most emphasize selected aspects of network evidence. Few systematically combine behavioral traffic patterns with visible service context. Their complementary value remains insufficiently explored, particularly when the reliability of each evidence view varies across activities and platforms.

\noindent\textbf{(3) Efficiency and deployment feasibility.} Reported execution settings include real-time laptop-based fingerprinting, gateway or controller deployments, and cloud middleboxes~\cite{flowprint2020,homesnitch2019,recon}. Online processing on these platforms does not establish native router feasibility. CATO~\cite{wan2025cato} jointly optimizes feature selection and packet depth to meet end-to-end latency and throughput budgets, while FlowWise~\cite{liu2026flowwise} cascades fast and slow models to keep pace with high flow-arrival rates. However, both rely on server-class infrastructure and tunable serving pipelines. It therefore remains unresolved whether traffic inference can accommodate heterogeneous cross-platform behavior while running natively on a resource-constrained intermediary such as a home router.
\section{System Design}
\label{sec:method}

\sysname{} is an end-to-end activity observation framework that transforms passive encrypted network traffic at the home router into unified, application-consumable activity and state histories across heterogeneous personal and IoT devices. Rather than acting as a standalone traffic classifier, \sysname{} bridges the gap between raw network packets and household applications by addressing three core design questions:

\begin{itemize}
    \item \textbf{Abstraction Gap.} How can predictions from heterogeneous inference pipelines be transformed into a common activity representation that supports application queries over sustained activities, discrete operations, and persistent device states? (\S\ref{sec:system-overview})
    \item \textbf{Segmentation Gap.} How can traffic be grouped into observations that retain context across concurrent flows for sustained activities while isolating intermittent operations from background traffic? (\S\ref{sec:task-adaptive-processing})
    \item \textbf{Behavior--Context Gap.} How can service context help interpret traffic patterns when services are shared across activities and context may be missing or ambiguous? (\S\ref{sec:activity-inference})
\end{itemize}

\subsection{System Overview}
\label{sec:system-overview}
\begin{figure}[t!]
\centering
\includegraphics[width=\linewidth]{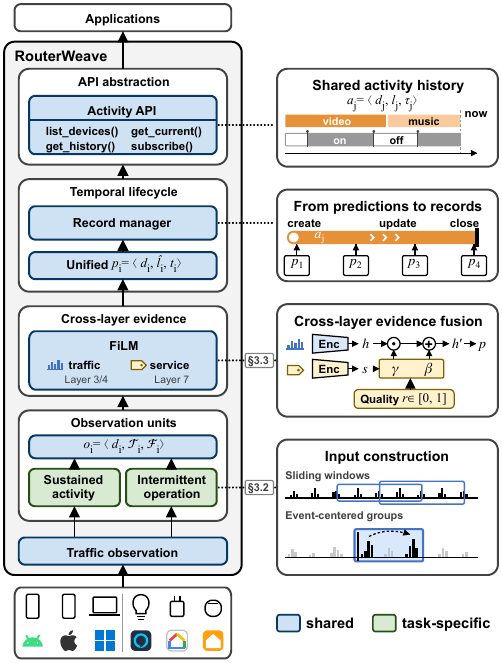}
\caption{Overview of \sysname{}. Shared traffic observation at the home router supports task-adaptive inference, while a common API exposes unified activity records across heterogeneous devices. Right: illustrations of key stages.}
\Description{Left, inside a \sysname{} frame and read from the bottom: home devices, each above its platform logo or app icon (an Android phone, an iPhone, a Windows laptop, an Alexa bulb, a Google Home plug and an Apple Home speaker), feed traffic observation; a task-adaptive processing group holds two green input paths, sustained activity and intermittent operation, both feeding a blue FiLM block over traffic and service features; a record management group holds the unified prediction p = (d, l, t), the record manager and the activity API with its four calls list\_devices, get\_current, get\_history and subscribe, which serves applications above the frame. Dotted leaders link four stages to framed examples on the right, with tags 3.3 and 3.2 on the fusion and input links: a shared activity history with an activity row and an on/off row; predictions creating, updating and closing an orange activity record; evidence fusion, where encoded traffic and service features combine through a scale and a shift; and input construction, with sliding windows over a sustained activity and the selected traffic of an operation.}
\label{fig:framework-layers}
\label{fig:architecture}
\label{fig:overview}
\end{figure}

Unlike prior traffic-analysis pipelines that only abstract raw packets into feature vectors for domain-specific classifiers, \sysname{} provides a complete observation layer on the home router. As illustrated in Figure~\ref{fig:framework-layers}, \sysname{} organizes traffic acquisition, task-adaptive processing, and record management into four abstraction tiers that progressively transform network packets into household activity APIs.

\noindent\textbf{Observation-unit abstraction: from raw packets to units that preserve activity evidence.}
At the bottom layer, traffic from all Wi-Fi and LAN devices passes through a shared observation point on the home router. However, a shared capture path does not imply a uniform segmentation rule: sustained activities, such as video streaming or web browsing on Android, iOS, and Windows, evolve across multiple concurrent connections and include quiet buffering intervals, whereas intermittent IoT operations, such as toggling a smart plug or appliance, manifest as brief, localized packet bursts amidst background heartbeats. Instead of imposing fixed time windows across all devices, \sysname{} abstracts raw packet streams into task-adaptive observation units of a common logical form:
\[
o_i = \langle d_i, \mathcal{T}_i, \mathcal{F}_i \rangle,
\]
where $d_i$ identifies the attributed source device, $\mathcal{T}_i$ denotes the temporal extent of the observation, and $\mathcal{F}_i$ encapsulates the selected packets and concurrent flows. By constructing recurring sliding windows across multiple flows for sustained activities and learned, event-centered packet groups that exclude background traffic for intermittent operations, this layer hides flow multiplicity and segmentation heterogeneity from upper stages (\S\ref{sec:task-adaptive-processing}).

\noindent\textbf{Cross-layer evidence abstraction: from protocol metadata to embeddings by evidence quality.}
Within each observation unit $o_i$, discriminative clues are scattered across multiple network layers. \sysname{} abstracts these protocol-level details into three components: (i) a behavioral traffic view $h$ derived from encrypted traffic, capturing L3/L4 packet timing, directional byte volumes, and concurrent connection dynamics; (ii) a passively visible service-context view $s$ capturing application-layer metadata from DNS queries, TLS handshakes, and local service discovery; and (iii) an evidence-quality indicator $r \in [0,1]$ that quantifies context ambiguity, DNS TTL aging, and flow coverage. Because behavioral statistics weaken during quiet media-buffering periods and service identifiers become ambiguous when multiple activities share CDN endpoints, \sysname{} uses feature-wise linear modulation (FiLM) to condition the behavioral embedding $h$ on the service context $s$, with $r$ controlling the strength of that adjustment (\S\ref{sec:activity-inference}). This layer handles protocol-specific details and uncertainty from missing metadata before classification.

\noindent\textbf{Temporal lifecycle abstraction: from heterogeneous predictions to unified activity records.}
Because sustained activities and discrete operations rely on different model families and inference cadences, \sysname{} decouples model execution from state management through a two-stage semantic contract. First, regardless of the underlying estimator, every task-adaptive pipeline emits outputs through a unified prediction interface:
\[
p_i = \langle d_i, \hat{\ell}_i, t_i \rangle,
\]
where $p_i$ is the $i$-th normalized prediction for device $d_i$ with inferred activity or operation label $\hat{\ell}_i$ at timestamp $t_i$. Second, the shared Record Manager consumes the stream of predictions $\{p_i\}$ and encapsulates temporal post-processing---such as state-continuity decoding, boundary refinement, and duplicate suppression---into three standard lifecycle operations (\texttt{create}, \texttt{update}, and \texttt{close}, as shown in Figure~\ref{fig:framework-layers}) to maintain unified activity records:
\[
a_j = \langle d_j, \ell_j, \tau_j \rangle,
\]
where $a_j$ denotes the $j$-th managed record in the shared activity history, $d_j$ identifies the source device, $\ell_j$ specifies the semantic activity or operation type, and $\tau_j$ defines its temporal support. The field $\tau_j$ unifies two distinct behavioral semantics within a single schema: for sustained personal-device activities, $\tau_j$ represents a continuous interval $[t_j^s,t_j^e]$ that is dynamically extended via \texttt{update} until the activity ends (\texttt{close}); for intermittent IoT operations, $\tau_j$ records a discrete transition instant $t_j$, allowing the Record Manager to maintain the device's resulting persistent operating state, such as ON or OFF, across subsequent periods of network silence.

\noindent\textbf{Application API abstraction: from shared history to application services.}
At the top of the framework, \sysname{} hides all packet capture, feature extraction, model inference, and temporal reconstruction logic behind four platform-independent API primitives (Figure~\ref{fig:framework-layers}):

\noindent\textbullet\enspace\texttt{list\_devices()} discovers the active devices observed on the local network and enumerates the activity or operation classes supported for each device.

\noindent\textbullet\enspace\texttt{get\_current()} returns a device's active ongoing activity or its currently maintained operating state.

\noindent\textbullet\enspace\texttt{get\_history()} retrieves historical activity records $\{a_j\}$ over a queried time horizon, enabling retrospective timeline reconstruction and duration aggregation.

\noindent\textbullet\enspace\texttt{subscribe()} provides a publish/subscribe stream that pushes real-time notifications to applications whenever an activity record $a_j$ is created, updated, or closed.

\subsection{Behavior-Preserving Traffic Input Construction}
\label{sec:task-adaptive-processing}

\providecommand{\windowfield}[1]{\textcolor{brown}{[#1]}}

Cross-platform heterogeneity affects which parts of continuous traffic should constitute an observation unit for activity recognition. A classifier receives useful evidence only if its input retains the traffic through which the target behavior is expressed~\cite{shaowang2025minimization}. For a sustained activity, that evidence may lie in the temporal evolution of several concurrent flows. On the other hand, for an intermittent operation, such a behavior may be concentrated in a few local exchanges. As a result, most past methods that use a single fixed traffic segmentation/preprocessing rule~\cite{flowprint2020,taylor2016appscanner} cannot preserve both structures: a short window can discard the history of a sustained activity or split an operation, while a long or indiscriminate window can dilute a brief operation with idle and background traffic.

\sysname{} thus treats input construction as a distinct, task-adaptive stage. Its goal is to avoid imposing identical windows across activity types and align the \emph{observation unit} with the temporal structure of the target behavior. Specifically, we propose adaptive traffic processing strategies for two distinct situations.

First, for sustained activities, \sysname{} adopts conventional sliding windows as temporal observation units. Each window retains recent traffic history, while successive windows capture how communication evolves over time, including bursts and quiet periods. As a result, such a strategy provides recurring observations of activities whose behavioral evidence persists beyond individual exchanges. Moreover, because sustained activities often access multiple resources and services concurrently, \sysname{} treats the flows within each window as a joint analysis unit rather than focusing on a single flow or an isolated packet burst. This construction preserves both aggregate temporal variation and the composition of concurrent connections, allowing downstream recognition to exploit temporal and cross-flow patterns together.

\begin{figure}[t!]
\centering
\includegraphics[width=\linewidth]{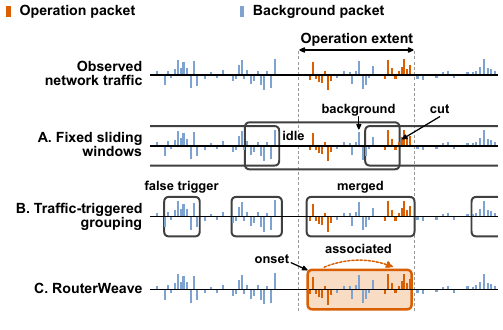}
\caption{(A) Fixed sliding windows cut the operation and add idle and background traffic, (B) traffic-triggered grouping merges background traffic into it, and (C) \sysname{} passes only the operation's packets, excluding the background between its exchanges.}
\label{fig:operation-input-localization}
\Description{Four aligned timelines repeat the same schematic packet trace: bursts of packet ticks of varying height above and below each axis, low-level blue background traffic, and one idle gap. The operation's two exchanges are orange and lie between two dashed extent lines. Dark-grey rounded boxes mark the inputs passed to the classifier. Equal, overlapping fixed windows include the idle gap and background, and one window boundary cuts the operation. Traffic-triggered grouping starts a false group on background traffic and merges the background between the exchanges into the operation. \sysname{}'s selection, tinted orange, runs from an onset mark at the operation's start to its last packet; a dotted orange arc associates the first exchange with the second, and the blue background between them is excluded.}
\end{figure}

On the other hand, for intermittent operations, \sysname{} instead constructs event-centered observation units that retain operation-related exchanges while excluding background traffic. Figure~\ref{fig:operation-input-localization} illustrates the limitations of two other conventional alternatives. The first option, \textit{fixed sliding windows}, can include substantial operation-independent background traffic, which obscures the evidence needed for recognition. Additionally, their fixed lengths and strides can also split an operation across windows, preventing the window from preserving the complete operation. The second alternative, \textit{traffic-triggered grouping}, aggregates sequences of uplink and downlink packets according to inter-packet gaps, but does not reliably isolate operations either. Burst characteristics vary across events, and background packets arriving during an operation can be merged into the same group, which similarly degrades downstream classification performance.

To this end, \sysname{} constructs observation units around \textbf{operational evidence} for intermittent operations rather than fixed time boundaries or traffic bursts alone. Using operation fingerprints extracted with PINGPONG~\cite{pingpong}, we train a small neural network with binary classification heads, which individually determine whether a packet is an \textit{operation start}, whether it is \textit{associated with a previous start}, or whether it is \textit{the end of an operation}. This packet-level membership prediction retains the initial operation-related burst and subsequent associated exchanges while excluding unrelated background packets. Consequently, temporally separated exchanges can contribute to the same observation even when unrelated packets arrive between them. The selected packets form an event-centered observation unit for downstream representation and recognition.

\subsection{Complementary Traffic Evidence Fusion and Representation Learning}
\label{sec:activity-inference}
\label{sec:complementary-evidence}

\providecommand{\featurefield}[1]{\textcolor{brown}{[#1]}}

A behavior-aligned observation unit determines the piece of traffic to observe, but not which aspects of that traffic are discriminative. As discussed in Section~\ref{sec:rw-traffic}, existing approaches often emphasize particular aspects of network evidence. One view consists of traffic statistics derived from encrypted traffic, including packet timing, sizes, directions, and the evolution of concurrent connections. These features describe how communication unfolds without relying on payload content. Another view comprises \emph{passively visible service context}, such as plaintext-derived DNS, handshake, and service-discovery evidence associated with the observation. Unfortunately, neither view is reliable or complete under every condition. Behavioral statistics can become weak or unstable during quiet intervals, within short observations, or when different implementations express the same activity through different communication patterns. Plaintext context can retain useful associations during such intervals, but it may be absent, shared across different activities, or ambiguously associated with individual flows. 

As a result, instead of relying on a single view, \sysname{} fuses information from both views so that each compensates for the other's limitations. The design problem is therefore not simply whether to add service context, but how to combine behavioral evidence with service context while accounting for their changing reliability.

\begin{figure}[t!]
\centering
\includegraphics[width=\linewidth]{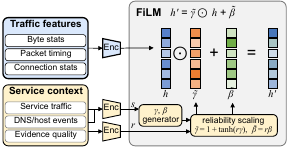}
\caption{Service-conditioned evidence fusion with FiLM. Traffic embedding $h$ is modulated by the reliability-scaled $\tilde{\gamma}$ and $\tilde{\beta}$.}
\label{fig:screen-time-film}
\Description{Traffic features (byte stats, packet timing, connection stats) pass through an encoder to the column h. In the service context, service traffic and DNS/host events pass through one encoder to s, and evidence quality passes through another encoder to r. A gamma, beta generator turns s into gamma and beta; a reliability scaling box, also given r, outputs gamma tilde equals 1 plus tanh of r gamma and beta tilde equals r beta. The FiLM panel shows columns h, gamma tilde, beta tilde and h prime, with h prime equal to gamma tilde times h plus beta tilde.}
\end{figure}

To combine these complementary views, \sysname{} adopts feature-wise linear modulation (FiLM)~\cite{perez2018}, a conditioning mechanism that uses one representation to modulate another through learned, feature-wise affine transformations. In our sustained-activity pipeline, the behavioral traffic representation remains the primary input to activity recognition, while service context conditions this representation according to the service environment in which the observed behavior occurs. This direct adjustment of the traffic embedding contributes to learning representations that align better with the corresponding network activities.

Specifically, as shown in Figure~\ref{fig:screen-time-film}, the traffic encoder transforms flow statistics such as packet sizes, arrival times and directional byte volumes into a behavioral embedding \(h\in\mathbb{R}^{256}\). The service-context encoder produces a context embedding \(s\) from passively visible DNS, TLS, and HTTP evidence. Two learned functions, \(\gamma(s)\) and \(\beta(s)\), map \(s\) to a scale and an offset for each dimension of \(h\). To prevent service context from dominating the embedding fusion, \sysname{} adopts an extra evidence-quality coefficient \(r\in[0,1]\), which measures the reliability of the underlying relationships between a service context and behavioral embedding. Overall, the modulated traffic representation is computed as
\[
\tilde{\gamma}(s,r)=1+\tanh\!\left(r\gamma(s)\right),\qquad
\tilde{\beta}(s,r)=r\beta(s),
\]
\[
h'=\operatorname{ReLU}\!\left(h\odot \tilde{\gamma}(s,r)+\tilde{\beta}(s,r)\right),
\]
where \(\odot\) denotes element-wise multiplication, and \(\tilde{\gamma}(s,r)\) and \(\tilde{\beta}(s,r)\) are the reliability-adjusted scale and offset, respectively. Intuitively, service context can amplify behavior dimensions that are informative in a particular service environment, suppress less relevant ones, or shift their values to obtain a fused activity representation. When service associations are weak or incomplete, \(r\) attenuates both modulation terms, preventing unreliable context from substantially altering the behavioral evidence. The resulting embedding \(h'\), rather than the unconditioned traffic representation, is passed to the activity classifier.
\FloatBarrier
\section{Evaluation}
\label{sec:experiments}
\begin{figure*}[t!]
\centering
\begin{tabular}{@{}c@{\hspace{10pt}}c@{\hspace{10pt}}c@{}}
\includegraphics[width=110pt]{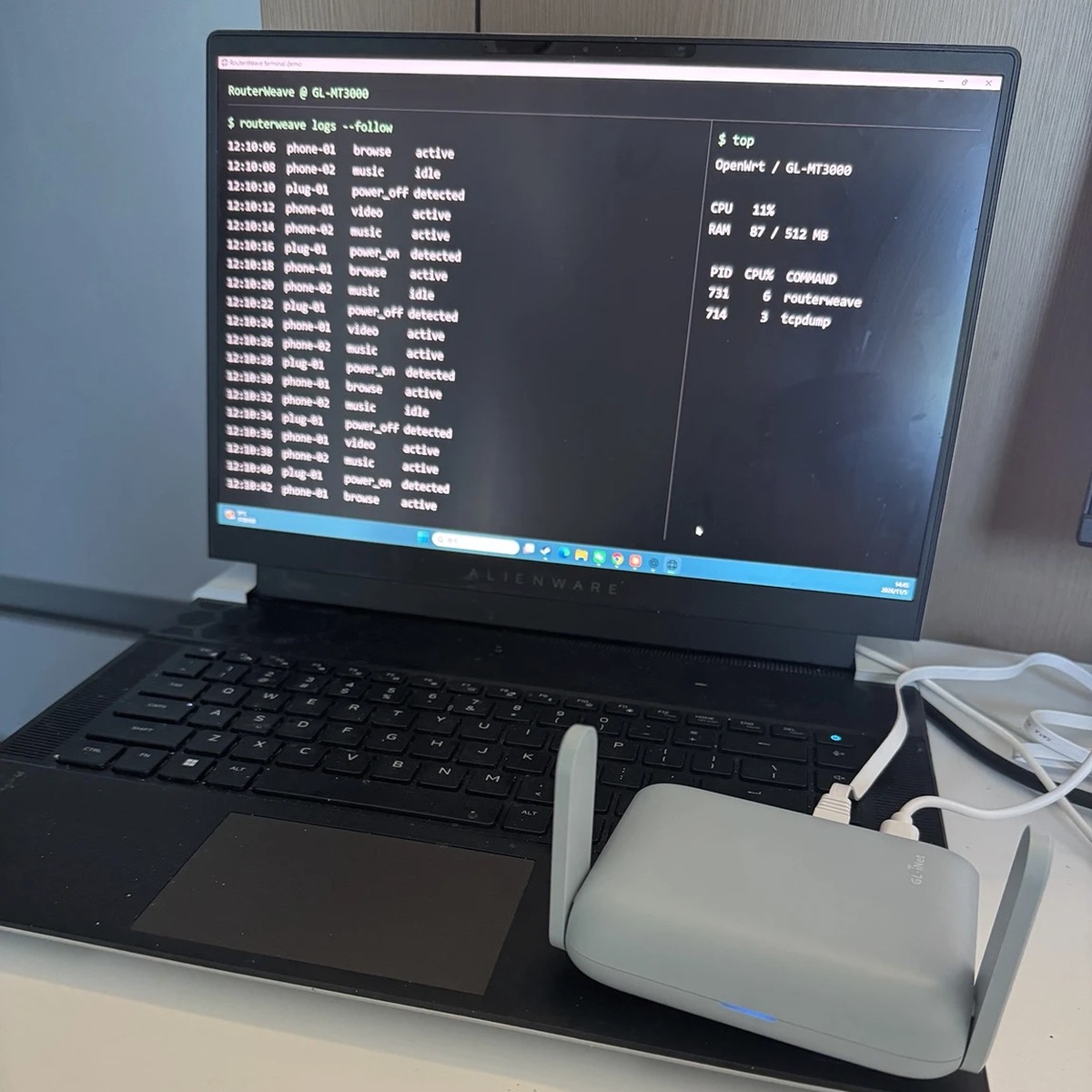} &
\includegraphics[width=110pt]{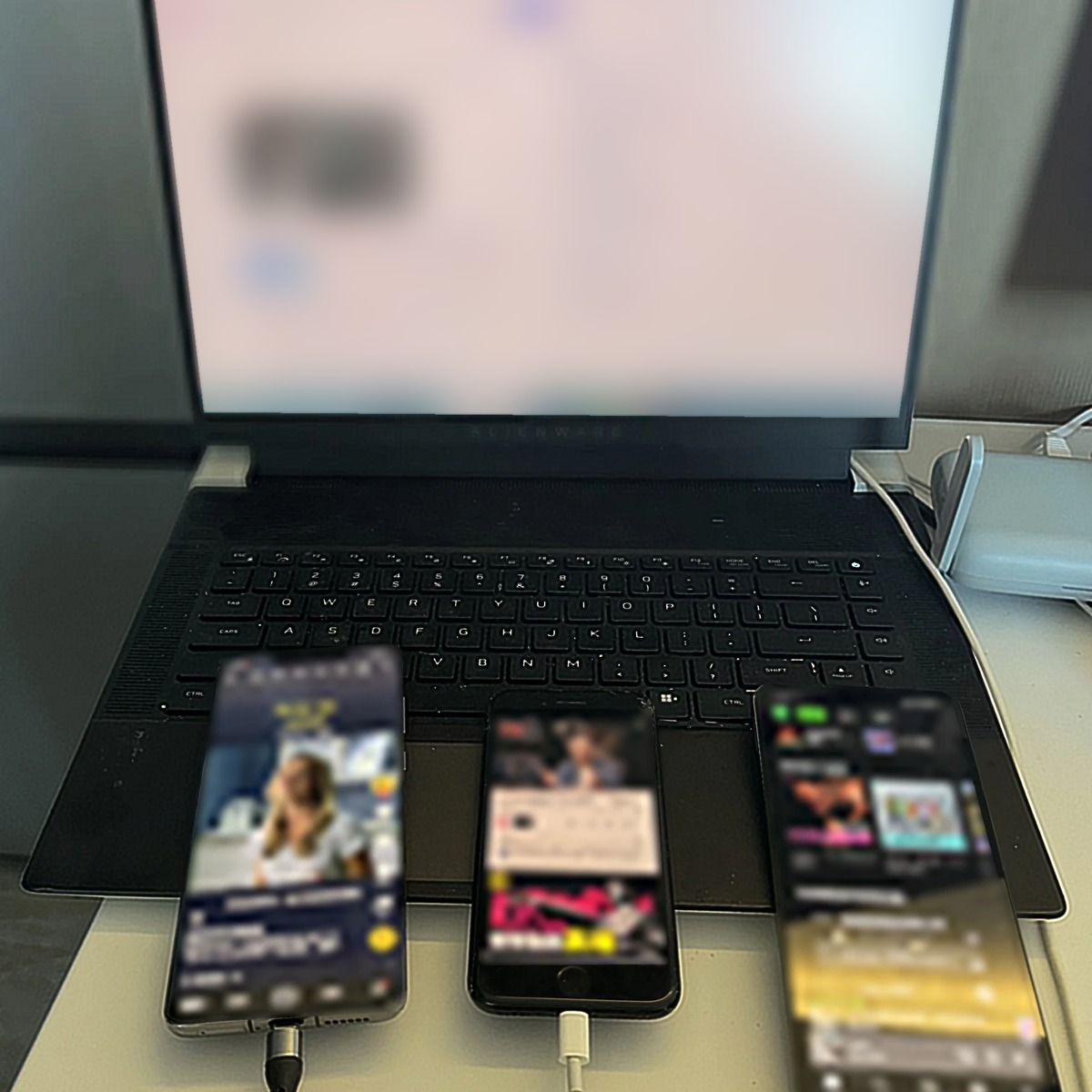} &
\includegraphics[width=110pt]{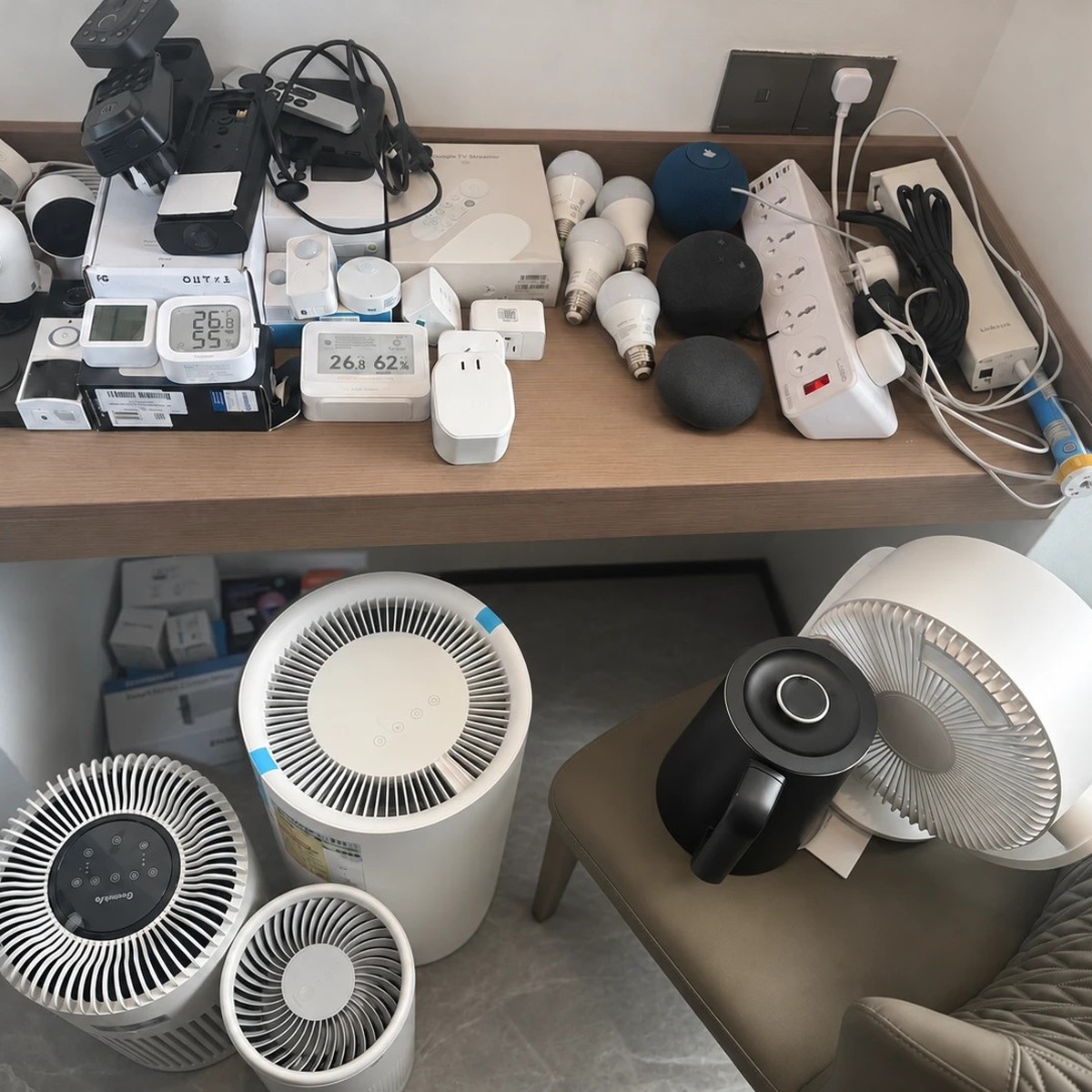}\\[2pt]
\small (a) Router deployment & \small (b) Cross-platform collection & \small (c) Wi-Fi IoT devices
\end{tabular}
\caption{Collection setup.}
\label{fig:collection-photos}
\Description{Three photographs. (a) A laptop shows the \sysname{} log, with device activities and operations, beside the router's resource monitor; a GL.iNet GL-MT3000 router sits on the laptop. (b) Three phones in front of a Windows laptop, all running applications during collection, with their screens blurred. (c) The Wi-Fi smart-home devices on a desk and the floor: cameras, doorbells, sensors, plugs, bulbs, smart speakers, a streaming box, air purifiers, a fan and a kettle.}
\end{figure*}
We evaluate \sysname{} through comparative recognition experiments and design analysis. In Section~\ref{sec:offline-analysis}, we compare \sysname{}'s recognition performance against two baseline approaches on four datasets, including both web/mobile application activities and IoT operations. In Section~\ref{sec:design-ablations}, we ablate and test the sensitivity of design choices in traffic input construction and evidence fusion / representation learning.
\subsection{Comparative Recognition Performance}
\label{sec:offline-analysis}
\subsubsection{Datasets and Evaluation Setup}
\paragraph{Customized Datasets.} We construct two datasets to evaluate activity recognition across operating systems and smart-home control platforms. The web/mobile application usage dataset covers 16 mobile applications and web services across Android, iOS, and Windows. These applications cover browsing, live streaming, long-form video, music, and short-form video running in both forms (foreground and background), using consistent activity labels across platforms. The IoT event dataset covers 15 device categories from 30 Wi-Fi devices on various platforms, including Alexa, Apple Home, and Google Home. During the collection, we repeatedly execute device operations, record their execution and state confirmation, and leave enough blank periods to finish potential long-tail communications before and after each action. In both collection processes, we use \texttt{tcpdump}~\cite{tcpdump} to capture traffic on a home router LAN bridge, and independently recorded annotations are aligned with packet traces based on their timestamps (Figure~\ref{fig:collection-photos}). The collections contain 78 hours of annotated application traffic and 5,040 verified IoT action records. Detailed collection procedures, dataset inventories, and label mappings are provided in Appendix~\ref{sec:implementation}.

\paragraph{External Benchmark Datasets.}
We additionally include two public datasets covering IoT activities and mobile application traffic. BehavIoT-Dataset, released with the BehavIoT method~\cite{behaviot2023}, provides labeled network traces of device operations collected from smart-home devices. ITC-Net-Blend-60~\cite{itcnetblend60} contains Android application traffic collected through real user interactions under different network conditions. 
Table~\ref{tab:recognition-datasets} summarizes the four datasets.
\begin{table}[t!]
\caption{Overview of the four datasets.}
\label{tab:recognition-datasets}
\centering
\footnotesize
\setlength{\tabcolsep}{3pt}
\renewcommand{\arraystretch}{1.15}
\begin{tabularx}{\linewidth}{@{}l>{\raggedright\arraybackslash}Xrrrr@{}}
\toprule
 & Platform / & Apps / & & & \\
Dataset & Ecosystem & Devices & Classes & Duration & Samples\\
\midrule
\multicolumn{6}{@{}l}{\textit{Application activities}}\\
\sysname{}-App & Android, iOS,\newline Windows & 16 & 6 & 78~h & 9,360\\
\addlinespace[2pt]
ITC-Net-Blend-60~\cite{itcnetblend60} & Android & 35 & 8 & 81~h & 10,020\\
\midrule
\multicolumn{6}{@{}l}{\textit{IoT operations }}\\
\sysname{}-IoT & Alexa,\newline Apple Home,\newline Google Home & 30 & 23 & 78.5~h & 5,040\\
\addlinespace[2pt]
BehavIoT-Dataset~\cite{behaviot2023} & Vendor\newline ecosystems & 30 & 18 & 60.3~h & 7,127\\
\bottomrule
\end{tabularx}
\end{table}

\paragraph{Baselines and Metrics.}
We compare \sysname{} with BehavIoT~\cite{behaviot2023} and FlowPrint~\cite{flowprint2020}, representing two established approaches to IoT behavior recognition and mobile application fingerprinting, respectively. BehavIoT employs per-device random forests over traffic statistics, while FlowPrint constructs fingerprints from correlations among network destinations. For each dataset, all methods are fitted on the same training partition and evaluated using identical test samples and prediction labels. We report macro-F1, supplemented by accuracy and per-platform results where applicable. Dataset-specific adaptations and training configurations are detailed in Appendix~\ref{sec:implementation}.

\subsubsection{Comparison Results}
Table~\ref{tab:four-dataset-baselines} reports macro-F1 on all four datasets. On the two datasets collected in this work, \sysname{} achieves 94.41\% macro-F1 for application activity recognition and 97.46\% for IoT operation recognition, exceeding the strongest baselines by 20.99 and 6.31 percentage points, respectively. Its performance remains competitive on independently collected traffic, reaching 87.72\% accuracy and 84.21\% macro-F1 on BehavIoT-Dataset.

In contrast, the baselines exhibit distinct strengths across activity types. FlowPrint performs better than BehavIoT on application recognition, whereas BehavIoT is more effective on IoT operations. This difference is consistent with their underlying mechanisms. FlowPrint fingerprints applications through correlations among network destinations, which may provide limited discrimination for IoT operations that share the same endpoints. BehavIoT uses packet sizes, timing, and flow statistics from localized traffic bursts to recognize device operations, but may miss evidence distributed across multiple connections and longer periods of application activity. These results highlight the limitations of applying a single recognition approach across heterogeneous activity types. In contrast, \sysname{} accommodates their different characteristics through task-adaptive processing, achieving the best performance on three of the four datasets.

\begin{table}[t!]
\centering
\caption{Macro-F1 (\%) comparison across four datasets. The best result for each dataset is shown in bold.}
\label{tab:four-dataset-baselines}
\small
\setlength{\tabcolsep}{2pt}
\renewcommand{\arraystretch}{1.15}
\begin{tabular*}{\linewidth}{@{\extracolsep{\fill}}lrrr@{}}
\toprule
Dataset & \sysname{} & BehavIoT & FlowPrint\\
\midrule
\sysname{}-App & \textbf{94.41} & 69.49 & 73.42\\
\addlinespace[2pt]
\sysname{}-IoT & \textbf{97.46} & 91.15 & 40.30\\
\addlinespace[2pt]
BehavIoT-Dataset & \textbf{84.21} & 70.79 & 37.34\\
\addlinespace[2pt]
ITC-Net-Blend-60 & 78.69 & 67.85 & \textbf{82.31}\\
\bottomrule
\end{tabular*}
\end{table}

\sysname{} performs less well on ITC-Net-Blend-60, achieving 79.90\% accuracy and 78.69\% macro-F1. One possible explanation is that the dataset captures natural user interactions without explicitly controlling or annotating the specific type of activities performed within each application. These interactions may be brief and highly variable, making it difficult for fixed-length observation windows to capture consistent behavioral patterns representative of the assigned application category.

\subsubsection{Efficiency and Resource Consumption}
\label{sec:router-efficiency}
\label{sec:framework-evaluation}
\label{sec:system-evaluation}
We further examine whether \sysname{}'s recognition gains incur substantial computational and resource costs on a resource-constrained home router. Table~\ref{tab:router-efficiency} compares macro-F1, output latency, and resource consumption on the GL-MT3000~\cite{glinetMT3000Specs}.

For application activities, \sysname{} achieves 20.99 percentage points higher macro-F1 than FlowPrint, at the cost of approximately 30 ms additional P95 prediction latency, while consuming less memory. This modest overhead suggests that improved recognition can be achieved without substantially delaying predictions. IoT recognition involves longer output delays because the system must accumulate and localize operation-related traffic before making a prediction. Nevertheless, \sysname{} achieves significantly higher macro-F1 and a lower P95 output delay than BehavIoT. Across both tasks, \sysname{} maintains small model and memory footprints, indicating that its task-adaptive recognition can operate within the limited resources of a home router.

\begin{table}[t!]
\caption{Macro-F1 gains and resource consumption measured on the GL-MT3000 home router.}
\label{tab:router-efficiency}
\centering
\small
\setlength{\tabcolsep}{2pt}
\renewcommand{\arraystretch}{1.15}
\begin{tabular*}{\linewidth}{@{\extracolsep{\fill}}llrrrr@{}}
\toprule
Task & Method & \shortstack{Macro-F1\\(\%) $\uparrow$} & \shortstack{P95\\latency $\downarrow$} & \shortstack{Peak RSS\\(MB) $\downarrow$} & \shortstack{Model size\\(MB) $\downarrow$}\\
\midrule
\multirow{2}{*}{Application} & \sysname{} & 94.41 & 41.08~ms & 6.99 & 2.50\\
 & FlowPrint & 73.42 & 11.62~ms & 19.77 & 2.83\\
\cmidrule{1-6}
\multirow{2}{*}{IoT} & \sysname{} & 97.46 & 4.77~s & 5.41 & 1.68\\
 & BehavIoT & 91.15 & 7.43~s & 0.88 & 0.48\\
\bottomrule
\end{tabular*}
\end{table}

We further examine whether \sysname{} can support concurrent activity monitoring across multiple devices in a typical household. Since a home router must also reserve resources for its operating system and networking functions, only a portion of its resources can be devoted to activity monitoring. For our router with 512 MB of RAM, we estimate a practical memory budget of approximately 100–150 MB for \sysname{}. Under these constraints, we evaluate workloads ranging from 2 to 54 logical devices. Even at 54 devices, the P95 processing time remains within the five-second update interval at approximately 4.8 s and peak process memory reaches 150 MB. These results demonstrate that \sysname{} can support 54 concurrent devices within the processing deadline and the upper end of the allocated memory budget. For comparison, a recent survey reports an average of 21 connected devices per US household~\cite{deloitte2023connectivity}. This measured capacity is more than twice the average household size, indicating that a single commodity router can meet the demands of household-wide activity monitoring.

\subsection{Design Analysis}
\label{sec:design-ablations}
\label{sec:component-analysis}
We next examine how two design choices contribute to these results: task-adaptive observation-unit construction and the combination of complementary evidence views.

\subsubsection{Task-Adaptive Observation-Unit Construction}
As discussed in Section~\ref{sec:task-adaptive-processing}, different activities require input-construction strategies suited to their temporal characteristics. Fixed sliding windows provide a natural way to observe sustained activities, but applying the same approach to intermittent operations may split related exchanges or mix them with unrelated background traffic. To this end, we compare the input construction with two conventional strategies: fixed sliding windows and traffic-triggered grouping based on configurable silence gaps. All three strategies use the same classifier type, training data, and event-matching criteria to isolate the effect of input construction.

\begin{figure}[t!]
\centering
\includegraphics[width=\linewidth]{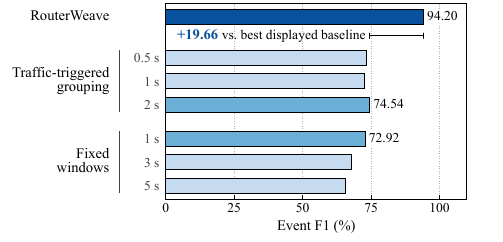}
\caption{Event recognition performance with different input construction methods. Time labels denote grouping thresholds or fixed-window lengths. The annotated gain is in percentage points.}
\label{fig:operation-input-results}
\Description{Horizontal bar chart of event F1. \sysname{}, at the top in dark blue, achieves 94.20 percent. Traffic-triggered grouping with silence gaps of 0.5, 1 and 2 seconds achieves 73.48, 72.56 and 74.54 percent. Fixed windows of 1, 3 and 5 seconds achieve 72.92, 67.79 and 65.74 percent. Medium blue marks the best displayed setting of each family, and a bracket marks \sysname{}'s 19.66-point margin over the best displayed baseline.}
\end{figure}

Figure~\ref{fig:operation-input-results} compares three input-construction strategies for intermittent-operation recognition across different temporal configurations. Fixed sliding windows achieve 65.74–72.92\% event F1 across the tested window lengths, while traffic-triggered grouping reaches 72.56–74.54\% under different silence thresholds. \sysname{} achieves 94.20\% F1, outperforming both alternatives. Fixed windows divide traffic according to predefined temporal boundaries, regardless of when operations occur, potentially splitting related exchanges or mixing them with background traffic. Traffic-triggered grouping alleviates this problem by aligning observation boundaries with actual communication. However, network activity does not necessarily indicate an operation: background traffic may trigger unnecessary observations, while pauses may separate exchanges belonging to the same operation.  \sysname{} addresses both issues by identifying potential operation onsets and associating subsequent exchanges based on operation-related evidence, thereby constructing more complete and focused inputs for recognition.

\subsubsection{Complementarity of Evidence Views}
As discussed in Section~\ref{sec:activity-inference}, \sysname{} combines behavioral traffic features with visible service context to capture complementary information about device activities. To examine the contribution of each evidence source, we compare independently trained traffic-only, service-only, and joint models using the same test windows and matched training configurations. The joint model achieves 94.76\% accuracy, compared with 76.16\% for traffic-only and 89.93\% for service-only recognition (Table~\ref{tab:evidence-views}). The improvement over both alternatives demonstrates the importance of combining behavioral traffic features with service context for more effective activity recognition.

\begin{table}[t!]
\caption{Recognition accuracy (\%) using joint traffic and service evidence or a single evidence view. Gain is the improvement over the corresponding single view, in percentage points.}
\label{tab:evidence-views}
\centering
\small
\setlength{\tabcolsep}{2pt}
\renewcommand{\arraystretch}{1.15}
\begin{tabular*}{\linewidth}{@{\extracolsep{\fill}}llrrr@{}}
\toprule
Test windows & Single view & Single & Joint & Gain\\
\midrule
\multirow{2}{*}{All} & Traffic only & 76.16 & \multirow{2}{*}{\textbf{94.76}} & +18.60\\
 & Service only & 89.93 & & +4.83\\
\addlinespace[2pt]
Buffered media & Traffic only & 60.32 & \textbf{96.83} & +36.51\\
Ambiguous services & Service only & 92.42 & \textbf{99.40} & +6.98\\
\bottomrule
\end{tabular*}
\end{table}

We dug into several cases to illustrate this complementarity. The first case occurs during sustained media activities where contents are buffered in advance for fluent display. The traffic pattern usually looks like an initial bursty downstream flow followed by prolonged periods of little network traffic. In such situations, recent traffic behavior offers limited evidence of the ongoing activity with 60.32\% traffic-only accuracy, while joint recognition reaches 96.83\% accuracy. Another situation arises when applications communicate with multiple services whose identities may not reliably reflect the application activity. For example, a service seemingly associated with live streaming may also appear during ordinary video playback, potentially leading to incorrect activity interpretation. Incorporating behavioral traffic features helps resolve this ambiguity, improving recognition accuracy from 92.42\% to 99.40\%.

Together, these results highlight the value of combining behavioral traffic features and service context in \sysname{}'s design. Each source can help resolve ambiguities left by the other. This complementarity allows \sysname{} to capture discriminative information that either source alone may miss, improving recognition across heterogeneous activities.

\section{Use Cases}
In this section, two realistic downstream case studies in research and industrial domains are presented to demonstrate \sysname{}'s feasibility and potential in the real world.
\label{sec:use-cases}
\subsection{Case Study I: Cross-Platform Screen Time Monitoring}
\label{sec:online-evaluation}

\begin{figure*}[t!]
\centering
\includegraphics[width=0.9\linewidth]{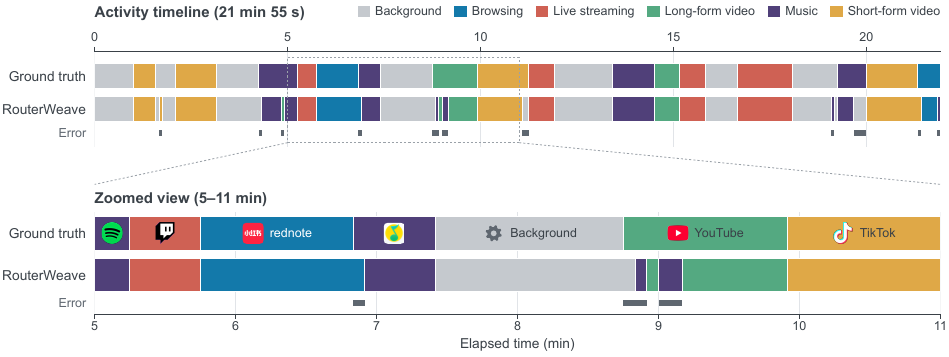}
\caption{Reconstruction of an activity timeline. Error strips mark label mismatches with ground truth. Icons identify the applications used in the ground-truth timeline.}
\label{fig:screen-time-monitoring-example}
\end{figure*}
We use cross-platform Screen Time monitoring as a realistic case study for evaluation of sustained activity monitoring.

\subsubsection{Application Scenario}
\label{sec:screen-time-use-case}
Digital well-being research has highlighted the importance of understanding people's activities across multiple digital devices and applications~\cite{rooksby2016personal,lyngs2024distraction}. Such understanding requires a comprehensive view of which devices people use, what activities they engage in, when these activities occur, and how long they last. However, this information remains fragmented across devices and platforms~\cite{appleScreenTime,googleFamilyLink,microsoftFamilySafety,samsungWellbeing}.

In this case study, we build a cross-platform Screen Time application on top of \sysname{}. Rather than collecting usage reports from individual platforms, the application uses \sysname{}'s common interface to examine and summarize usage across devices. For historical analysis, it uses \texttt{get\_history()} to retrieve activity records over selected periods and calculate usage durations for individual devices. For ongoing monitoring, it uses \texttt{subscribe()} to receive activity updates as they occur and \texttt{get\_current()} to retrieve the latest activity estimates.

\subsubsection{Evaluation Setup}
\label{sec:experimental-setup}
\label{sec:screen-time-setup}

To evaluate both retrospective and live monitoring, we use 600 synthetic activity timelines/sequences using the previously recorded activity and background traffic. We first form 500 randomized activity sequences by randomly sampling applications with various usage durations (sampled from 10–30, 30–90, and 120–300 second ranges), and placing them in order. Such randomly generated activity sequences aim at mimicking potential users' behaviors on their mobile/computer usage, such as frequently switching doom-scrolling apps (or even platforms). Due to the balanced sampling strategy among apps and platforms, the randomness also stress-tests the capability of \sysname{} on cross-platform activity recognition. The remaining 100 activity sequences follow more realistic timing patterns derived from real-user traces collected by previous research~\cite{aliannejadi2021lsapp}. We substitute the apps and the corresponding traffic on the timelines with our previously collected segments chronologically. The substitution depends on the categories of the respective activities, e.g., long/short video, music, etc. As traffic arrives, \sysname{} continuously infers device activities and maintains their records through successive updates at five-second intervals.

\subsubsection{Results}
\textit{Retrospective Usage Analysis.} Figure~\ref{fig:screen-time-monitoring-example} shows a representative reconstructed history alongside its ground truth. Successive activities and background periods are recovered, while the remaining errors affect activity attribution: part of a long-form video interval, for instance, is classified as music. Such errors may preserve the total active duration while distorting the time attributed to individual activities.

Across the 600 constructed timelines, averaging approximately 19 minutes each, \sysname{} correctly labels 92.15\% of the evaluated time and estimates total active duration with a mean absolute error of 24.84 seconds per timeline.

Table~\ref{tab:screen-time-overall-results}(a) further examines reconstruction quality across operating systems. \sysname{} achieves at least 88.64\% time-weighted accuracy and 74.48\% temporal \textbf{intersection over union} (IoU) on each platform, demonstrating that the same downstream analysis can recover activity histories from heterogeneous platforms with consistent activity categories and temporal representations.

Table~\ref{tab:screen-time-overall-results}(b) examines reconstruction performance under different activity durations. Even for timelines with short (10--30-second) activity segments and frequent switches, \sysname{} achieves 86.10\% time-weighted accuracy and 75.29\% temporal IoU, with further improvements for longer activities. These results show that \sysname{} can maintain effective activity reconstruction across a range of usage patterns, including frequent activity changes.

\begin{table}[t!]
\centering
\caption{Timeline reconstruction by (a) platform and (b) switching frequency. Accuracy is time-weighted; IoU denotes temporal intersection over union. Ranges in (b) indicate individual activity durations.}
\label{tab:screen-time-overall-results}
\footnotesize
\setlength{\tabcolsep}{3pt}
\renewcommand{\arraystretch}{1.15}
\begin{tabular*}{\linewidth}{@{\extracolsep{\fill}}lrrrr@{}}
\toprule
\multicolumn{5}{@{}l}{(a) Platform}\\
\midrule
 & Android & iOS & Windows & Overall\\
\midrule
Accuracy & 91.48\% & 88.64\% & 96.55\% & 92.15\%\\
IoU & 75.10\% & 74.48\% & 90.99\% & 79.32\%\\
\bottomrule
\end{tabular*}
\par\medskip
\begin{tabular*}{\linewidth}{@{\extracolsep{\fill}}lrrr@{}}
\toprule
\multicolumn{4}{@{}l}{(b) Switching frequency}\\
\midrule
 & frequent/10--30(s) & moderate/30--90(s) & slow/120--300(s)\\
\midrule
Accuracy & 86.10\% & 94.16\% & 96.06\%\\
IoU & 75.29\% & 85.62\% & 82.43\%\\
\bottomrule
\end{tabular*}
\end{table}

Together, these findings demonstrate the feasibility of \sysname{}'s unified activity-observation framework for cross-platform Screen Time monitoring. The application can reconstruct device-level activity histories and estimate usage durations across different operating systems, even under frequent activity changes, using the information provided through a common interface without relying on platform-specific usage reports.

\paragraph{Live Activity Monitoring.}
We further evaluate \sysname{}'s ability to maintain an accurate and timely view of activities as they unfold. \sysname{} correctly represents the ongoing activity for 88.44\% of the evaluated time. When activities change, it successfully reports 79.74\% of the transitions, with an average delay of 6.61 seconds for matched changes. These results demonstrate \sysname{}'s ability to continuously track ongoing activities and respond to changes with low latency.

Together with the retrospective reconstruction results, they show that \sysname{}'s unified framework supports two complementary capabilities: continuous monitoring of current activities and reconstruction of historical usage across heterogeneous platforms.

\subsection{Case Study II: Appliance Energy Estimation}
\label{sec:electricity-use}

\newcommand{\applOpsCorrect}{49}
\newcommand{\applOpsLogged}{55}
\newcommand{\applErrors}{six}
\newcommand{\applKettleCorrect}{10}
\newcommand{\applKettleLogged}{12}
\newcommand{\applPurifierErrors}{three}
\newcommand{\applPurifierMissed}{two}
\newcommand{\applPurifierWrong}{one}
\newcommand{\applPurifierMissedHours}{6.6}
\newcommand{\applPurifierWrongHours}{20.7}
\newcommand{\applPurifierWrongWh}{122.7}
\newcommand{\applPurifierAgreement}{71.46}
\newcommand{\applAgreement}{92.84}
\newcommand{\applKettleCycleWh}{101.3}
\newcommand{\applEstimatedWh}{8133.5}
\newcommand{\applReferenceWh}{8497.5}
\newcommand{\applEnergyError}{4.28}
\newcommand{\applPurifierEnergyError}{31.93}
\newcommand{\applKettleEnergyError}{16.67}
\newcommand{\applOperationRows}{Air purifier & 11 & 8 & 2 & 1\\ Dehumidifier & 18 & 17 & 1 & 0\\ Circulation fan & 14 & 14 & 0 & 0\\ Electric kettle & 12 & 10 & 2 & 0\\ }
\newcommand{\applOperationTotal}{Overall & 55 & 49 & 5 & 1\\}
\newcommand{\applEnergyRows}{Air purifier & 506.5 & 344.7 & 31.93\\ Dehumidifier & 6201.0 & 6201.2 & $<$0.01\\ Circulation fan & 575.1 & 575.1 & $<$0.01\\ Electric kettle & 1215.0 & 1012.5 & 16.67\\ }
\newcommand{\applEnergyTotal}{Total & 8497.5 & 8133.5 & 4.28\\}
\newcommand{\applDehumidifierCorrect}{17}
\newcommand{\applDehumidifierLogged}{18}
\newcommand{\applKettleMissed}{two}
\newcommand{\applKettleMissedWh}{202.5}
We use smart appliance energy estimation as the case study for the evaluation of intermittent activity monitoring.
\subsubsection{Application Scenario}
Household electricity measurements reveal how much energy is consumed, but provide limited insight into how individual appliances contribute to that consumption. Understanding when appliances operate and how long they remain active can help residents become aware of daily electricity use and identify potential energy waste~\cite{saha2014energylens}. However, appliance-level measurements often require built-in sensing capabilities or additional hardware such as power meters.

In this case study, we build an appliance energy estimation application on top of \sysname{}. Rather than measuring each appliance directly, the application uses \sysname{}'s common interface to retrieve appliance operating states and their transitions, and estimates the electricity usage within the active periods by referring to the devices' supplied average power values.

\subsubsection{Evaluation Setup}
\label{sec:iot-setup}
\sysname{} is deployed on the home router to continuously monitor four network-connected appliances: an air purifier, a dehumidifier, a circulation fan, and an electric kettle. Each appliance is observed over a four-day period, during which multiple ON/OFF operations are performed at varying intervals to reflect everyday usage and are logged as ground-truth references.

The downstream application accesses these states through \sysname{}'s \texttt{get\_history()}, \texttt{get\_current()}, and \texttt{subscribe()} APIs. Electricity consumption is estimated by combining the inferred operating durations with device-specific power values measured using external plug-in power meters. %
For the kettle, each ON operation initiates a preset three-minute heating cycle, after which the kettle automatically switches off. The evaluation compares the inferred operating histories with the reference logs and the estimated electricity consumption with reference calculations based on the same power assumptions.

\subsubsection{Results}

We evaluate the application along the chain from recovered operations to maintained appliance states and the resulting electricity estimates.

\paragraph{Operation Recovery.} \sysname{} correctly recovers 49 of the 55 recorded operations within the ±10-second tolerance (Table~\ref{tab:appliance-operations}). These observations provide the state transitions needed to track appliance operation over time, without requiring continuous communication from each device.

\begin{table}[t!]
\centering
\caption{\sysname{} recovers \applOpsCorrect{} of the \applOpsLogged{} logged operations within $\pm10$~s.}
\label{tab:appliance-operations}
\small
\setlength{\tabcolsep}{3pt}
\begin{tabular*}{\linewidth}{@{\extracolsep{\fill}}lrrrr@{}}
\toprule
Appliance & Logged & Correct & Missed & Wrong direction\\
\midrule
\applOperationRows
\midrule
\applOperationTotal
\bottomrule
\end{tabular*}
\end{table}

\paragraph{State Maintenance.} Figure 7 demonstrates how brief network-observable operations can support state tracking over much longer periods. \sysname{} maintains operating states even when no further operation-related traffic is observed, achieving 92.84\% agreement with the reference over the evaluated time. Most discrepancies arise from missed or incorrectly identified transitions. For example, the air purifier's ON and OFF operations share the same observable traffic fingerprint, making their directions difficult to distinguish. Consequently, one incorrectly assigned transition causes its maintained state to remain inconsistent with the reference for 20.7 hours. This illustrates how the reliability of long-term state maintenance depends on correctly interpreting the underlying operations.

\begin{figure}[t!]
\centering
\includegraphics[width=1.05\linewidth]{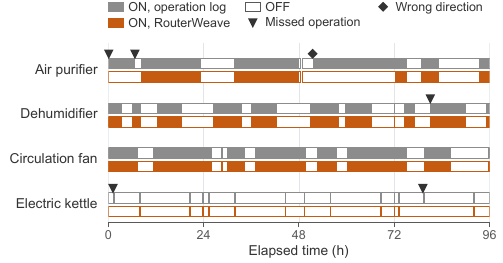}
\caption{\sysname{}'s maintained states match the operation logs except after the air purifier's missed and wrong-direction operations. Markers above the logs show every operation error over the four days.}
\label{fig:appliance-state}
\Description{Four pairs of bars on each appliance's own elapsed time from 0 to 96 hours: the logged state in grey above the state maintained by \sysname{} in orange, OFF in white. Dark markers above the logged bars mark operation errors: two missed operations and one wrong direction for the air purifier near the start and at about 51 hours, one missed dehumidifier operation at about 81 hours, and two missed kettle cycles near the start and at about 79 hours. The air purifier's \sysname{} bar is OFF after its errors while the log is ON; the other pairs match.}
\vspace{1em}
\end{figure}

\paragraph{Electricity Estimation.}
\label{sec:energy-evaluation}
The downstream application uses these maintained states, together with measured appliance power consumption, to estimate electricity usage. Across the four appliances, \sysname{} estimates 8133.5 Wh, compared with 8497.5 Wh derived from the smart plug logs, a difference of 4.28\% (Table~\ref{tab:appliance-energy}). The estimates closely match the reference for the dehumidifier and circulation fan, while larger deviations occur for the air purifier and kettle. The relatively small aggregate difference is largely driven by the accurately tracked dehumidifier, which accounts for most of the total consumption.

Together, these results demonstrate that \sysname{}'s activity information is not limited to the moment an operation is recognized. Through unified activity management, brief network observations can be transformed into persistent operating-state information, enabling downstream applications to monitor device behavior over time and derive quantitative estimates such as household electricity consumption.

\begin{table}[t!]
\centering
\caption{Estimated energy stays within \applEnergyError\% of the log-based total; the air purifier and kettle account for nearly all of the difference. Both use the same measured power values.}
\label{tab:appliance-energy}
\small
\setlength{\tabcolsep}{3pt}
\begin{tabular*}{\linewidth}{@{\extracolsep{\fill}}lrrr@{}}
\toprule
Appliance & Log-based (Wh) & \sysname{} (Wh) & Error (\%)\\
\midrule
\applEnergyRows
\midrule
\applEnergyTotal
\bottomrule
\end{tabular*}
\end{table}
\FloatBarrier
\section{Discussion}
We further discuss how \sysname{} could evolve beyond its current scope to provide a more comprehensive understanding of household activities.
\subsection{Toward Resident-Centric Activity Observation}
\sysname{} provides a unified view of activities across heterogeneous devices, opening opportunities to understand how individual residents interact with their device ecosystems. In practice, the relationship between residents and devices is not one-to-one: a resident may switch between multiple devices, while shared devices may be used by different household members. Consequently, understanding a resident's behavior requires connecting activities across devices rather than examining each device's activity history.

One promising direction is to infer resident--activity associations by combining device-specific information with relationships among observed activities. Device ownership, access permissions, and online presence can provide initial cues, while individual usage patterns and temporal relationships across devices may help identify the residents associated with shared-device activities. These associations could be continuously refined as new activities are observed, allowing \sysname{}'s unified records to support resident-level activity timelines across personal and shared devices. This would enable applications to examine individual screen time, daily routines, and interactions with the smart-home environment from a resident-centric perspective.

\subsection{Extending Activity Observation Across Home Networks}
\sysname{} currently uses a home router as a shared observation point, with our evaluation focusing on Wi-Fi-connected devices. More complex smart homes, however, may involve other communication technologies such as Zigbee~\cite{csaZigbee} and Thread~\cite{threadGroup}. Interestingly, activities on these networks may still leave recognizable traces in the IP traffic of intermediary devices. For example, prior studies have demonstrated that operations of Zigbee devices can be inferred from encrypted traffic generated by their hubs~\cite{pingpong,chatterhub2021}.

These observations suggest that \sysname{} could extend its activity coverage beyond Wi-Fi devices by incorporating traffic evidence associated with other communication technologies into its router-based recognition framework. This would allow a broader range of smart-home activities to be observed and managed through the same unified framework.

\subsection{Granularity of Activity Observation}
\sysname{} provides a unified view of activities across heterogeneous devices, but the granularity of behavioral information remains constrained by what network traffic reveals. While many activities can be reliably recognized, finer-grained operations may produce nearly identical traffic patterns. For example, turning a light on or off and adjusting its color may all generate short encrypted exchanges with similar packet sizes and timing, making it difficult to distinguish these actions.

One possible direction is to exploit additional evidence beyond the immediate operation. For example, devices in different operating states may exhibit distinct background traffic patterns, providing clues about their operating states. Moreover, associated controllers, such as smartphones running control apps or smart-home hubs, may generate additional communication that helps differentiate these operations. Correlating such evidence across time and devices could enable \sysname{} to recognize finer-grained operations.
\subsection{Incremental Support for New Devices and Activities}
\sysname{} currently supports a predefined set of devices and activities, while real-world smart homes continuously evolve as new devices and services are introduced. Extending activity coverage currently requires collecting labeled traffic and training corresponding recognition models, which could become costly as the number of supported devices grows.

A promising direction is to enable lightweight activity enrollment by transferring knowledge from previously supported devices. Recent advances in few-shot IoT traffic classification suggest that learned traffic representations can be adapted using limited labeled examples~\cite{zhao2026muffin}. For new activity types, class-incremental learning techniques~\cite{cerasuolo2024memento} could further support the introduction of new activities while preserving previously learned capabilities. Beyond expanding coverage, maintaining recognition accuracy also requires adapting to changes in existing devices. Studies of concept drift in residential IoT traffic~\cite{pashamokhtari2023drift} suggest opportunities for adapting models to firmware and software updates~\cite{sivanathan2020behavior}.

\section{Conclusion and Future Work}

We presented \sysname{}, a router-native framework that turns passive network traffic into unified activity histories across personal and IoT devices. Through task-adaptive input construction, cross-layer evidence fusion, and stateful lifecycle management, \sysname{} delivers high recognition accuracy within the tight memory and latency budgets of a commodity router. Evaluations across four datasets and two end-to-end case studies in cross-platform Screen Time tracking and appliance energy estimation show that existing home gateways can serve as shared household sensing infrastructure without endpoint software or vendor APIs.

Looking forward, unified router-side observation unlocks fundamental directions in both privacy and longitudinal health sensing. On one hand, the ability to reconstruct cross-device timelines from encrypted traffic motivates deeper investigation into household privacy leakage and mitigation. On the other hand, \sysname{} provides a non-invasive foundation for long-term well-being monitoring, such as linking rapid cross-platform attention switching to behavioral assessment for ADHD and tracking subtle drifts in daily appliance routines to monitor Alzheimer's disease progression.

\section*{Ethics Statement} 

All custom network traffic in this study was collected on dedicated, researcher-operated testbeds and devices within an isolated laboratory network. No external human participants or private household traffic were monitored. All PIIs are removed to protect privacy.
\input{main.bbl}

\appendix
\section{Recognition Models}
\label{sec:appendix-models}

\subsection{Application Activity}
\sysname{} processes 30-second observations represented as one-second traffic bins, concurrent connection features, and packet-header features (Table~\ref{tab:appendix-feature-dim}). A traffic encoder produces a 256-dimensional embedding, modulated through FiLM using service context and evidence quality. Four classification experts, each with a $256\!\rightarrow\!128\!\rightarrow\!6$ architecture, produce probabilities for five activity categories and background. Training balances platforms, classes, and recordings, with platform-grouped PCGrad.

\begin{table}[!htbp]
\caption{Input features for application activity recognition.}
\label{tab:appendix-feature-dim}
\centering\small
\begin{tabularx}{\linewidth}{@{}>{\raggedright\arraybackslash}p{0.23\linewidth}>{\raggedright\arraybackslash}X@{}}
\toprule
\textbf{Feature group} & \textbf{Features and representation} \\
\midrule
Device traffic & Directional packet counts, byte volumes, TCP packet counts, packet-size and interarrival-time distributions; 30 one-second bins.\\
Concurrent connections & 16 static attributes and two directional traffic time series per connection.\\
Packet headers & Directional packet and byte counts, TCP/UDP shares, TCP payload and flag statistics, TTL distributions, and advertised TCP-window distributions; 40 features.\\
Service context & 3 global service counts, 324 grouped service statistics, and 44 hostname/DNS event features.\\
Evidence quality & 9 quality indicators for the observed service evidence.\\
\bottomrule
\end{tabularx}
\end{table}

\subsection{IoT Operations}
\sysname{} first localizes operation-related packets and then classifies the selected sequence. Each physical device uses one localizer across its control platforms, with a 32-unit encoder and a bidirectional GRU producing start, membership, and end predictions. A second bidirectional GRU encodes the selected packets for operation classification. FiLM conditioning uses 42 preceding-traffic features and a known-device embedding, while three quality features regulate modulation (Table~\ref{tab:appendix-iot-features}). Classifier training settings are summarized in Table~\ref{tab:appendix-classifier-training}.

\begin{table}[!htbp]
\caption{Input features for IoT operation recognition.}
\label{tab:appendix-iot-features}
\centering\small
\begin{tabularx}{\linewidth}{@{}>{\raggedright\arraybackslash}p{0.23\linewidth}>{\raggedright\arraybackslash}X@{}}
\toprule
\textbf{Feature group} & \textbf{Features and use} \\
\midrule
Packet sequence & Packet size, direction, and time since the preceding device and connection packets; input to the localizer.\\
Burst structure & Packet position and elapsed time within the burst; input to the localizer.\\
Candidate relations & Relative time and position, and shared connection and burst membership with the proposed start; used for membership and end prediction.\\
Selected traffic & The packet sequence retained by the localizer; input to the operation classifier.\\
Preceding traffic & 42 features describing traffic preceding the candidate operation; used for FiLM conditioning.\\
Device identity & An embedding of the known physical device; used for FiLM conditioning.\\
Evidence quality & Three quality features controlling modulation strength.\\
\bottomrule
\end{tabularx}
\end{table}

\begin{table}[!htbp]
\caption{Classifier training configuration.}
\label{tab:appendix-classifier-training}
\centering\small
\begin{tabular}{@{}lcc@{}}
\toprule
\textbf{Setting} & \textbf{Application} & \textbf{IoT operation} \\
\midrule
Optimizer & AdamW & AdamW \\
Learning rate & $10^{-3}$ & $10^{-3}$ \\
Weight decay & $10^{-3}$ & $10^{-4}$ \\
Batch size & 256 & 128 \\
Epochs & 7 (final refit) & 40 \\
\bottomrule
\end{tabular}
\end{table}

\section{Comparative Recognition Experiments}
\label{sec:implementation}

\subsection{Datasets and Scoring Units}
\label{sec:appendix-data}

\paragraph{\sysname{}-App.}
Mobile activity intervals began when the controller verified the requested action. Android UID and socket traces and iOS process captures identified the app. Windows labels followed the foreground process. Fourteen background-only recordings supplied the background label. In 48 development recordings, previously used apps retained their background state while the foreground activity supplied the label.

\begin{table*}[!tp]
\begin{minipage}[t]{0.54\textwidth}
\centering
\caption{\sysname{}-App services by activity.}
\label{tab:screen-dataset}
\small
\setlength{\tabcolsep}{3pt}
\renewcommand{\arraystretch}{1.12}
\begin{tabularx}{\linewidth}{@{}l>{\raggedright\arraybackslash}Xc@{}}
\toprule
Activity & Applications & Platforms\\
\midrule
Browsing & Baidu, CSDN, Weibo, Xiaohongshu & \multirow{5}{*}{\shortstack[c]{Android\\iOS\\Windows}}\\
Live streaming & Douyu, Huya, Twitch & \\
Long-form video & Bilibili, iQIYI, YouTube & \\
Music & NetEase Cloud Music, QQ Music, Spotify & \\
Short-form video & Douyin, Kuaishou, TikTok & \\
\bottomrule
\end{tabularx}
\end{minipage}\hfill
\begin{minipage}[t]{0.43\textwidth}
\centering
\caption{\sysname{}-App split by recording. Test recordings were disjoint from fitting but had previously been examined during development.}
\label{tab:screen-dataset-details}
\small\setlength{\tabcolsep}{3pt}
\begin{tabular}{@{}lrrr@{}}
\toprule
Platform & \shortstack{Development\\captures} & \shortstack{Test\\captures} & \shortstack{Test\\windows}\\
\midrule
Android & 64 & 15 & 900\\
iOS & 53 & 14 & 852\\
Windows & 39 & 10 & 672\\
\midrule
\textbf{Total} & \textbf{156} & \textbf{39} & \textbf{2,424}\\
\bottomrule
\end{tabular}
\end{minipage}
\end{table*}

Only complete 30-second windows within a labeled interval were eligible. Training windows advanced by five seconds; validation and test windows did not overlap.

\paragraph{\sysname{}-IoT.}
Table~\ref{tab:iot-offline-dataset} lists the collection inventory. The classifier is scored on a subset of these operations, with opposite directions merged where appropriate. For example, raising and lowering brightness both map to \emph{light/brightness}. The 19 operation outputs cover power and fan speed for purifiers and circulation fans; power for dehumidifiers, kettles, and plugs; brightness, color, and power for lights; playback and volume for speakers and media players; curtain movement; lock operation; doorbell presses and live view; and periodic sensor reports. Background is a separate output.

\begin{table*}[!tp]
\centering
\caption{Wi-Fi IoT devices, control platforms, and operations.}
\label{tab:iot-offline-dataset}
\small
\setlength{\tabcolsep}{2.5pt}
\renewcommand{\arraystretch}{1.02}
\newcommand{\iotentry}[1]{\begin{tabular}[t]{@{}l@{}}#1\end{tabular}}
\begin{tabular}{@{}llll@{}}
\toprule
Device type & Devices & Platforms & Operations / events\\
\midrule
Lights &
\iotentry{Amazon Basics A19\\Meross MSL120DA\\SmartLife/Tuya E27 bulb} &
\iotentry{Alexa\\Alexa, Apple Home, Google Home\\Alexa, Google Home} &
\iotentry{Switch on/off\\Increase/decrease brightness\\Change color}\\
\addlinespace[4pt]
Plugs &
\iotentry{Doiting CozyLife-Switch-L1\\Meross MSS110\\Quectel Matter plug} &
\iotentry{Alexa, Apple Home\\Alexa, Apple Home, Google Home\\Alexa, Apple Home} &
Switch on/off\\
\addlinespace[4pt]
Locks &
\iotentry{TCL D1\\ULTRALOQ Bolt Fingerprint} &
\iotentry{Alexa\\Alexa, Apple Home, Google Home} &
Lock/unlock\\
\addlinespace[4pt]
Curtains &
\iotentry{Zemismart ZM82A\\Zemismart ZM25W01} &
\iotentry{Alexa, Apple Home\\Alexa, Google Home} &
Open/close\\
\addlinespace[4pt]
Air purifiers &
\iotentry{Govee H7122\\Meross MAP100} &
\iotentry{Alexa, Google Home\\Alexa, Apple Home} &
\iotentry{Switch on/off\\Increase/decrease fan speed}\\
\addlinespace[4pt]
Fan & Xiaomi Desktop Circulation Fan & Alexa &
\iotentry{Switch on/off\\Increase/decrease fan speed}\\
\addlinespace[4pt]
Dehumidifier & Xiaomi Dehumidifier Lite & Alexa & Switch on/off\\
\addlinespace[4pt]
Kettle & Wi-Fi cordless kettle & Alexa & Switch on/off\\
\addlinespace[4pt]
Speakers &
\iotentry{Amazon Echo Dot (4th gen.)\\Apple HomePod mini\\Google Nest Mini} &
\iotentry{Alexa\\Apple Home\\Google Home} &
\iotentry{Play/pause\\Increase/decrease volume\\Query voice assistant}\\
\addlinespace[4pt]
Media players &
\iotentry{Apple TV\\Google TV} &
\iotentry{Apple Home\\Google Home} &
\iotentry{Switch on/off\\Play/pause\\Increase/decrease volume}\\
\addlinespace[4pt]
Cameras &
\iotentry{Aqara Camera Hub G2H Pro\\Google Nest Cam Indoor\\TP-Link Tapo C100} &
\iotentry{Apple Home\\Alexa, Google Home\\Alexa, Google Home} &
Start/stop live view\\
\addlinespace[4pt]
Doorbells &
\iotentry{Aqara Video Doorbell G4\\TP-Link Tapo D210\\Google Nest Doorbell} &
\iotentry{Alexa, Apple Home\\Alexa, Google Home\\Alexa, Google Home} &
\iotentry{Start/stop live view\\Press the doorbell button}\\
\addlinespace[4pt]
Motion sensor & SmartLife PIR sensor & Alexa, Google Home & Motion detected/cleared\\
\addlinespace[4pt]
Presence sensor & Meross presence sensor & Google Home & Presence detected/cleared\\
\addlinespace[4pt]
Environmental sensors &
\iotentry{Govee H5103\\INKBIRD IBS-TH5-WIFI} &
\iotentry{---\\---} &
Report periodically\\
\bottomrule
\end{tabular}
\end{table*}

Section~\ref{sec:offline-analysis} scores 1,022 detector-generated candidates matched to reference operations. Accuracy and macro-F1 are averaged over fits with seeds 42, 43, and 44. Of the 74 evaluation capture batches, 39 contributed distinct samples to fitting. Missed operations and unmatched predictions enter the event-level comparison in Section~\ref{sec:design-ablations}.

\paragraph{ITC-Net-Blend-60.}
Whole captures were assigned to training, validation, and testing (399, 129, and 149 captures). They yielded 5,885, 2,031, and 2,104 nonempty 30-second windows, respectively, including the final partial window. Each window inherits the application's dominant-purpose label (Table~\ref{tab:appendix-itc}). For example, a call in a messaging app is labeled messaging. The other 25 apps were excluded.

\begin{table}[!htbp]
\caption{Application-purpose labels used for ITC-Net-Blend-60.}
\label{tab:appendix-itc}
\centering\small
\begin{tabularx}{\linewidth}{@{}l>{\raggedright\arraybackslash}X@{}}
\toprule
Label & Applications\\
\midrule
Browsing & Fidibo, Taghche, Goodreads, Google Chrome, Firefox Browser, Divar, Digikala, Shaypur, Torob\\
Music & Radio Javan, Castbox, Spotify\\
Video & Telewebion, YouTube, Aparat, Filimo\\
Gaming & Clash of Clans, Football Strike, Mencherz, Quiz of Kings, Mafioso\\
Messaging & WhatsApp Messenger, WhatsApp Business, Telegram, iGap, Soroush Plus Messenger, Eitaa\\
Email & Gmail, Microsoft Outlook\\
Navigation & Balad, Neshan, Google Maps, Waze\\
Conferencing & Google Meet, Adobe Connect\\
\bottomrule
\end{tabularx}
\end{table}

\paragraph{BehavIoT-Dataset.}
Methods receive the activity clip boundaries. The partition contains 4,223 training, 1,408 validation, and 1,409 test clips. Nested clips from a common capture component stay in the same partition.

\subsection{Comparative Experiment Settings}

All methods use the same training partition, test samples, and prediction labels within each dataset. We report macro-F1, supplemented by accuracy and per-platform results where applicable. BehavIoT recognizes activities from burst statistics using one-versus-rest random forests. FlowPrint matches destination and certificate fingerprints using Jaccard similarity. Table~\ref{tab:appendix-baseline-settings} summarizes their adaptations.

\begin{table}[!htbp]
\caption{Baseline implementation settings.}
\label{tab:appendix-baseline-settings}
\centering\small
\begin{tabularx}{\linewidth}{@{}>{\raggedright\arraybackslash}p{0.23\linewidth}>{\raggedright\arraybackslash}X@{}}
\toprule
\textbf{Component} & \textbf{Setting} \\
\midrule
BehavIoT input & Split flows at a one-second silence gap; compute 22 statistics per burst.\\
Model scope & One set of forests per physical IoT device or operating system for \sysname{}-App.\\
Training samples & Idle/background bursts supply negatives when available; at most 300,000 application bursts per platform.\\
Sample prediction & Aggregate one-versus-rest forest decisions within each evaluation sample.\\
Adaptation & Numerical burst features; no hostname filtering or periodic-traffic subtraction.\\
\addlinespace
FlowPrint input & Destination and certificate fingerprints.\\
Matching & Global fingerprint matching by Jaccard similarity.\\
Sample prediction & Majority flow label within each evaluation sample.\\
\bottomrule
\end{tabularx}
\end{table}

FlowPrint matches globally, whereas the \sysname{} IoT classifier restricts its output labels using the known device. Capture-level splitting allows destinations to recur between training and testing.

\subsection{Router Experiment Settings}
\label{sec:appendix-router}

\paragraph{Runtime Measurement.}
All experiments are conducted on a GL-MT3000 router. Latency experiments preserve packet and timer timestamps while skipping idle waiting periods; concurrent-stream experiments use paced replay.

\begin{table}[!htbp]
\caption{Router environment and workload settings.}
\label{tab:appendix-router-settings}
\centering\small
\begin{tabularx}{\linewidth}{@{}>{\raggedright\arraybackslash}p{0.23\linewidth}>{\raggedright\arraybackslash}X@{}}
\toprule
\textbf{Component} & \textbf{Setting} \\
\midrule
Hardware & GL-MT3000; two CPU cores and 512~MB RAM.\\
Runtime & Native AArch64 code under Linux 6.6.110.\\
Application workload & Ten minutes each of Android QQ Music, iOS Bilibili, and Windows YouTube; upper-median packet-rate recording per platform.\\
Application repeats & 115 updates per trace; 30 repetitions, giving 10,350 timings in total.\\
IoT workload & 50 recordings, of which 46 contain a reference onset; 30 repetitions.\\
IoT closure & \sysname{} uses a two-second quiet period and five-second expiry in latency tests.\\
Concurrent load & 2--54 logical devices; equal numbers of Windows and Google Nest Mini trace replicas, sharing weights with separate stream state.\\
Run duration & 30-second warm-up followed by 20 five-second rounds per concurrent load.\\
\bottomrule
\end{tabularx}
\end{table}

\paragraph{Latency.}
Application latency is measured from window cutoff to prediction output, including parsing and feature extraction but excluding observation time, activity reconstruction, and API delivery. IoT latency is measured from operation onset to the first overlapping positive prediction, regardless of class, and is reported only for matched operations. BehavIoT latency ends at its burst decision, before window closure and voting.

\paragraph{Resource Usage.}
Peak RSS includes preloaded models, while model size includes workload-specific weights and indices. CPU utilization is normalized by the router's two CPU cores.

\section{Use-Case Evaluation}
\label{sec:appendix-cases}

\subsection{Screen Time Monitoring}
\paragraph{Timeline sampling.}
Randomized timelines last 15--30 minutes on a five-second grid. Each draws a background sampling weight uniformly from $[0.05,0.35]$ and splits the remainder among foreground classes by a symmetric Dirichlet draw with concentration 2 per class. At each switch the preceding label is excluded and the weights renormalized. Excerpts do not overlap within a timeline. The 500 layouts reuse 130 development recordings; 50 are development layouts and 37 were used to fit the temporal heads.

The user-guided timelines use 100 nonoverlapping LSApp sequences from 49 Android users. Original spans last at least 30 seconds. The 854 seconds of unknown gaps remain unscored; 40,062 seconds have assigned foreground labels. Timing comes from LSApp and packet excerpts from 68 controlled Android recordings (43 training, 12 development, 13 test). Both cohorts use separate flow namespaces per excerpt and carry forward the service association established before the crop in its original recording.

\paragraph{Record publication.}
The immediate GRU combines 18 log probabilities from 5-, 10-, and 30-second observations with 26 traffic, service-availability, and probability-consistency features. The delayed GRU pairs current and target vectors with two availability flags. Each GRU has 16 hidden units. Both were fitted on 80 development timelines with the ten-epoch classifier and frozen when used with the seven-epoch classifier. Early confirmation checks pending labels at zero, five, and ten seconds; at fifteen seconds the delayed decision is used. Publication follows interval order.

\paragraph{Scoring.}
Scoring excludes the 60-second warm-up. Accuracy pools correctly labeled seconds; temporal IoU averages over classes with nonempty union within a timeline, then equally over timelines. Active-duration MAE averages absolute foreground-duration error per timeline. Platform results combine both cohorts; duration profiles cover only the randomized cohort. Live-monitoring accuracy scores the immediate label of the just-completed five-second interval on the 500 randomized timelines. A published change must match both old and new labels, follow the reference transition within 30 seconds, and precede or coincide with the next reference transition. Recall counts all reference changes. Delay averages replay publication time minus reference time for one-to-one matches, excluding capture, computation, and transport.

\subsection{Appliance Energy Estimation}
\paragraph{Operation references.}
We continuously collected network traffic and corresponding operation logs for all four appliances over four days. The logs record the timing and type of each appliance operation and serve as the reference for evaluating \sysname{}'s inferred operations.

\paragraph{State durations.}
For the purifier, dehumidifier, and fan, both reference and estimated state histories are initialized to the state opposite the first logged switch. Each state persists until an opposing operation or the end of the recording. Reference states are updated using logged operations, while estimated states are updated using \sysname{}'s inferred operations. Each kettle start contributes a 180-second operating interval, truncated at the end of the recording.

\paragraph{Power coefficients.}
Reference and estimated energy consumption are calculated by integrating the power values in Table~\ref{tab:appendix-power} over their respective state durations. Reference durations are derived from the operation logs, whereas estimated durations are derived from inferred operations. Thus, the reference represents calculated energy consumption rather than an independent measurement of cumulative electricity use. The resulting reference energy consumption is 506.5~Wh for the air purifier, 6,201.0~Wh for the dehumidifier, 575.1~Wh for the circulation fan, and 1,215.0~Wh for the kettle, totaling 8,497.5~Wh.

\begin{table}[!htbp]
\caption{Power used to integrate logged and inferred on/off durations.}
\label{tab:appendix-power}
\centering\small
\begin{tabular}{@{}lrr@{}}
\toprule
Appliance & On (W) & Standby (W)\\
\midrule
Air purifier & 6.94 & 1.02\\
Dehumidifier & 117.00 & 1.08\\
Circulation fan & 7.08 & 3.48\\
Electric kettle & 2,025 & 0\\
\bottomrule
\end{tabular}
\end{table}
\end{document}